\documentclass[12pt]{article}

\pdfoutput=1
\usepackage{scicite}
\usepackage{authblk}
\usepackage{times}
\usepackage{amsmath}
\usepackage{amssymb}
\usepackage{float}
\usepackage{placeins}
\usepackage{graphicx}
\usepackage{braket}
\allowdisplaybreaks

\usepackage{xcolor}
\definecolor{red}{rgb}{1,0.,0}

\newenvironment{sciabstract}{%
\begin{quote} \bf}
{\end{quote}}

\begin{document} 


\title{Simulations of Subatomic Many-Body Physics on a Quantum Frequency Processor\footnote{This manuscript has been authored by UT-Battelle, LLC under Contract No. DE-AC05-00OR22725 with the U.S. Department of Energy. The United States Government retains and the publisher, by accepting the article for publication, acknowledges that the United States Government retains a non-exclusive, paid-up, irrevocable, world-wide license to publish or reproduce the published form of this manuscript, or allow others to do so, for United States Government purposes. The Department of Energy will provide public access to these results of federally sponsored research in accordance with the DOE Public Access Plan. (http://energy.gov/downloads/doe-public-access-plan).}}

\author[1]{Hsuan-Hao Lu}
\author[2]{Natalie Klco}
\author[3]{Joseph M. Lukens}
\author[3,6]{Titus D. Morris}
\author[4]{Aaina Bansal}
\author[5]{Andreas Ekstr\"om}
\author[6,4]{Gaute Hagen}
\author[4,6]{Thomas Papenbrock}
\author[1]{Andrew M. Weiner}
\author[2]{Martin J. Savage}
\author[3,$\dagger$]{Pavel Lougovski}

\affil[1]{School of Electrical and Computer Engineering and Purdue Quantum Center, Purdue University, West Lafayette, Indiana 47907, USA}
\affil[2]{Institute for Nuclear Theory, University of Washington, Seattle, WA 98195-1550, USA}
\affil[3]{Quantum Information Science Group, Computational Sciences and Engineering Division, Oak Ridge National Laboratory, Oak Ridge, Tennessee 37831, USA}
\affil[4]{Department of Physics and Astronomy, University of Tennessee, Knoxville, TN 37996, USA}
\affil[5]{Department of Physics, Chalmers University of Technology, SE-412 96 G{\"o}teborg, Sweden}
\affil[6]{Physics Division, Oak Ridge National Laboratory,
Oak Ridge, TN 37831, USA}
\affil[$\dagger$]{To whom correspondence should be addressed; E-mail: lougovskip@ornl.gov}

\date{}


\baselineskip24pt

\maketitle

\begin{sciabstract}
Simulating complex many-body quantum phenomena is a major scientific impetus behind the development of quantum computing, and a range of technologies are being explored to address such systems. 
We present the results of  the largest photonics-based simulation to date, 
applied in the context of subatomic physics. 
Using an all-optical quantum frequency processor, the ground-state energies of light nuclei including the triton ($^3$H), $^{3}$He, and the alpha particle ($^{4}$He) are computed. 
Complementing these calculations
and utilizing a 68-dimensional Hilbert space,
our photonic simulator is used to perform sub-nucleon calculations of the two-body and three-body forces between heavy mesons in the Schwinger model. This work is a first step in simulating subatomic many-body physics on quantum frequency processors---augmenting classical computations that
bridge scales from quarks to nuclei.
\end{sciabstract}

\paragraph*{Introduction}
Photonics is at the forefront of experimental quantum computing, as evidenced by pioneering demonstrations of the variational quantum eigensolver (VQE) algorithm~\cite{lanyon2010towards,peruzzo2014variational,SantagatiWAVES2018},   of molecular vibronic spectra and dynamics simulations~\cite{Huh2015,Sparrow2018}, and of experimental Hamiltonian learning~\cite{Wang2017}. It offers a versatile platform to process quantum information with low noise in a multitude of encodings, ranging from spatial or polarization degrees of freedom ~\cite{Kok2007,crespi2011integratedCNOTpolarization},  to temporal modes~\cite{Humphreys2013, Ansari2018}. Rapid progress in integrating optical components on-chip~\cite{Harris2017,Wang2018,qiang2018} is paving the way to large-scale spatial-encoding-based photonic quantum processors. Other encodings, however, also provide a path to scalable quantum architectures. For example, frequency encoding---routinely used in fiber optics to multiplex information transmission and processing---has been adapted for scalable quantum computing~\cite{Lukens 2017}. A single fiber can support thousands of frequency modes that can be manipulated in a massively-parallel fashion at the single-photon level. This particular framework for photonic quantum computing relies on qubits encoded in narrow frequency bins, where quantum gates are based on standard telecommunication equipment: electro-optic phase modulators (EOMs) and Fourier-transform pulse shapers~\cite{Lukens 2017}. A variety of basic quantum functionalities have recently been demonstrated experimentally in this approach, in the form of a quantum frequency processor (QFP)~\cite{Lu2018a,Lu2018b,Lu2018c}. 

Solving quantum many-body systems, whose resource requirements scale exponentially with the number of particles, is an area in which quantum devices are anticipated to provide a quantum advantage over classical computation.
Recently, quantum many-body problems in chemistry, condensed matter, and subatomic physics have been addressed with quantum computing using two-to-six superconducting qubits, for example  Refs.~\cite{omalley2016,kandala2017hardware,Dumitrescu2018,Klco:2018kyo,Lamm:2018siq}, and up to tens of trapped ions, for example Refs.~\cite{Martinez:2016yna,Zhang2017,Kokail:2018}. 

Here we report the first application of a QFP to photonic simulations of many-body subatomic systems. 
Our results involve experimental energy minimization in Hilbert spaces of up to 68 dimensions, and represent the largest implementation of nuclei and lattice quantum field theories on photonic devices to date.
In particular, using an effective field theory (EFT) description, we experimentally implement the VQE algorithm to calculate the binding energies of the atomic nuclei $^3$H, $^{3}$He, and $^{4}$He. Further, for the first time, we employ VQE to determine the effective interaction potential between composite particles directly from an underlying lattice quantum gauge field theory, the Schwinger model. This serves as an important demonstration of how EFTs themselves can be both implemented and determined from first principles by means of quantum simulations.

\begin{figure}[!ht]
\centering
\includegraphics[width=0.95\textwidth]{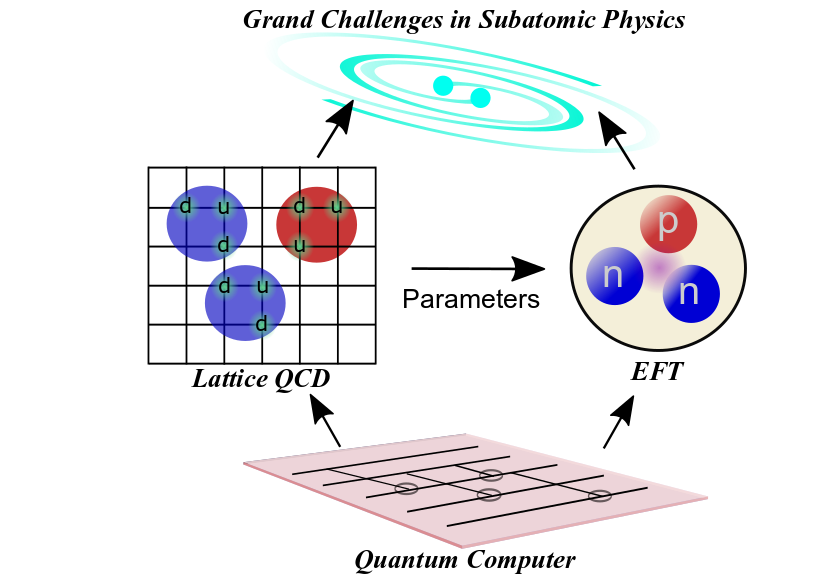}
\caption{
Quantum simulation for subatomic physics. 
Ideally, quantum simulation applied to both QCD (left-side) and EFT (right-side) will enable high-precision predictions of static and dynamic properties of nuclei and nuclear matter.
EFT parameters may be determined from experiment, or by a complementary program of classical and quantum simulation. 
}
\label{fig1}
\end{figure}

A major goal in nuclear physics research is to tie the EFT descriptions  of nuclear matter and heavy nuclei to their microscopic origin, quantum chromodynamics (QCD), through numerical calculations with lattice QCD. 
Important steps are being taken toward this objective~\cite{Yamazaki:2009ua,Beane:2011iw,Barnea:2013uqa,Beane:2012ey,Beane:2012vq,Yamazaki:2012hi,Inoue:2014ipa,Yamazaki:2015asa,Kirscher:2015yda,Contessi:2017rww,Bansal:2017pwn,Iritani:2018zbt}.  
A hierarchy of EFT models~\cite{bedaque2002,epelbaum2009,machleidt2011} is used
to describe heavier nuclei~\cite{navratil2009,barrett2013,hagen2015,morris2018b}, 
and lattice QCD calculations have been used to constrain EFT parameters over a range of unphysical quark masses~\cite{Beane:2012ey,Barnea:2013uqa,Kirscher:2015yda,Contessi:2017rww,Bansal:2017pwn}, 
with the goal of
extending the reach of lattice QCD calculations to heavier nuclei, rare isotopes, and dense matter in astrophysical environments.
However, such microscopic descriptions are computationally challenging at present for all but the lightest nuclei and hypernuclei~\cite{Yamazaki:2009ua,Beane:2011iw,Beane:2012ey}
due to signal-to-noise problems~\cite{Parisi:1983ae,Lepage:1989hd,Beane:2009kya}.
Augmenting classical calculations with their quantum counterparts~\cite{Ovrum:2007a,Jordan:2011ne,Jordan:2011ci,Marshall:2015mna} offers an analogous roadmap for quantum-enabled subatomic physics simulations as depicted in Fig.~\ref{fig1}.  
At the EFT level, a subatomic system can be simulated as a collection of nucleons with  EFT parameters input from
experimental data or \emph{ab initio} calculations.
At a microscopic level, Minkowski-space quantum simulations are proposed to compute these parameters directly from lattice QCD.
In this article, 
we take the first steps to 
meeting this Grand Challenge.
Using a photonic QFP, we compute the ground-state energies of several light nuclei using experimentally-determined EFT parameters, and in the lattice Schwinger model---a prototypical theory sharing important features with QCD---we show how EFT parameters can be informed by simulating the effective interaction potential between two and three composite  particles.

\paragraph*{Quantum Frequency Processor (QFP)}
For implementing quantum simulations, we utilize our previously-developed QFP: a photonic device that processes quantum information encoded in a comb of equispaced narrow-band frequency bins, described by operators $c^{\dagger}_n$ ($c_n$) for $n \in \mathbb{Z}$ that create (annihilate) a photon in a mode centered at the frequency $\omega_n=\omega_0+n\Delta\omega$, where $\Delta\omega$ is the frequency bin spacing and $\omega_0$ is an offset~\cite{Olislager2010, Lukens2017}. Mathematically, the QFP is described by a unitary mode transformation matrix $V$ that connects input $c_n^{(\mathrm{in})}$ and output $c_m^{(\mathrm{out})}$ modes, so that $c_m^{(\mathrm{out})} = \sum\limits_{n} V_{mn}c_n^{(\mathrm{in})}$. An arbitrary transform $V$ can be implemented by interleaving pulse shapers and EOMs~\cite{Lukens2017}, and recent experiments have demonstrated high-fidelity single-qubit gates~\cite{Lu2018a,Lu2018b} and a two-qubit controlled-NOT~\cite{Lu2018c}. 

\begin{figure}[!ht]
\centering
\includegraphics[width=0.6\textwidth]{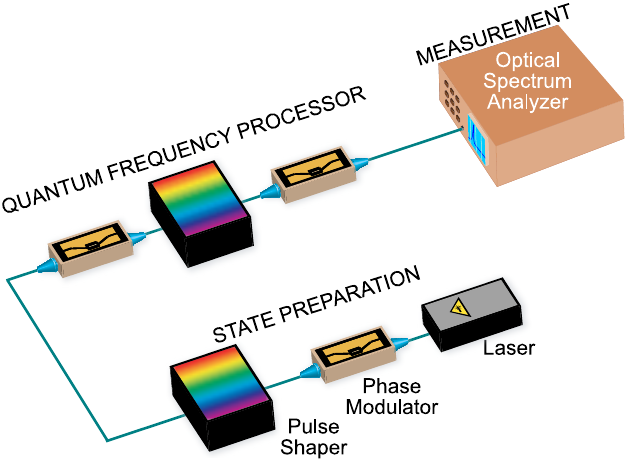}
\caption{Experimental setup including our all-optical quantum frequency processor.}
\label{expSetup}
\end{figure}

Figure~\ref{expSetup} shows the experimental setup for the all-optical QFP we implement here. The input state preparation, frequency operations, and the final energy measurements can all be realized with off-the-shelf fiber-optic components, including EOMs (EOSpace), Fourier-transform pulse shapers (Finisar), and an optical spectrum analyzer (OSA; Yokogawa). The capability of transmitting optical information within a single-mode fiber from generation to detection facilitates parallel computations in a low-noise fashion.

As discussed below, for many-body Hamiltonians projected onto single-particle sub-spaces a variational wavefunction can be mapped onto a mode-entangled state of a single photon, so that the state preparation procedure in the VQE algorithm amounts to the generation of a coherent frequency comb. However, more complicated (e.g., multi-photon) entangled photonic states could be employed as well, modifying only the ``State Preparation'' portion of the apparatus in Fig.~\ref{expSetup}.
By working with multiple photons in the QFP, qubit degrees of freedom can be identified with photon occupations of frequency-bin pairs.  For example, 10 frequency bins, discussed below, can be mapped onto 5 qubits with a 5-photon input state. Such a mapping and the ability to implement a universal gate set
endows the QFP with similar quantum capabilities and scaling as other digital quantum devices. Note that the QFP utilized in this work can in principle support upto 33 qubits~\cite{Lu2018a}.  Scaling this hardware to larger numbers of qubits will require further engineering in order to build multiple multi-photon sources and reduce loss in the system.

Specifically, the source we utilize for the state preparation is a wavelength-division multiplexing transmitter (WDM transmitter; Ortel) possessing four continuous-wave (CW) laser modules emitting single-tone optical signals at 192.0, 192.4, 192.8, and 193.2 THz. We combine all four signals with a $4 \times 1$ fiber coupler and send them through an EOM driven at 25 GHz, which creates four parallel frequency combs at the output (each of which contains $\sim$10 frequency bins at 25-GHz spacing) with a total of 40 available comb lines. This allows us to implement up to eight parallel Hadamard gates (50/50 beamsplitters): two comb lines per gate plus four comb lines for guardbands (i.e., two gates per each 10-line subcomb) to prevent cross-contamination during the subsequent frequency beamsplitting operations in the QFP (photons from one mode pair jumping over to an adjacent pair)~\cite{Lu2018a}. 
We then choose the best five beamsplitters for the subsequent operation, excluding those with higher imbalance in reflectivity and transmissivity. Note that despite these calibration measures, we still observe the residual non-uniform performance across different gates stemming from the noise background of the input light source, and the optical path length change in the QFP due to temperature drifts in long-term measurements. Finally, filtering out extra input frequencies, as well as manipulating amplitude and phase for all five remaining frequency-bin pairs, is realized by a pulse shaper immediately following the first EOM.

\paragraph*{Mapping Problems onto QFP}
In all quantum simulations here, our starting point is a second-quantized Hamiltonian $\mathcal{H}_{SQ}$ which, depending on the problem, contains one-, two-, and three-particle terms written as products of fermionic creation and annihilation operators. Our goal for these Hamiltonians is to compute the smallest eigenvalue using the QFP hardware. A scalable path to this goal has been outlined in the literature in the form of the VQE algorithm~\cite{mcclean2016}. There, each fermionic operator in $\mathcal{H}_{SQ}$ is mapped onto a set of qubits such that fermionic commutation relationships are preserved. As a result $\mathcal{H}_{SQ}$ is mapped onto $\tilde{\mathcal{H}}_{SQ}$ which is a sum of strings of Pauli operators.  Then quantum hardware is used to prepare a variational trial quantum state of qubits $|\Psi = \Psi(\theta_1,\cdots,\theta_M)\rangle$ in the form of a parameterized quantum circuit with $M$ parameters.
Subsequently, the expectation value of the Hamiltonian in the state $|\Psi\rangle$, $\langle\Psi|\tilde{\mathcal{H}}_{SQ}|\Psi\rangle$,  is computed by repeating the state preparation and energy measurement multiple times. A classical computer calculates the direction in the parameter space and new parameter values $\{\theta^{\prime}_1,\cdots,\theta^{\prime}_M\}$ that yield a lower energy value. The energy calculation is then repeated on quantum hardware with the updated trial state $|\Psi(\theta^{\prime}_1,\cdots,\theta^{\prime}_M)\rangle$ until a (local) minimum of the energy is obtained. 

For all pre-error-corrected quantum hardware---of which the QFP is an example---the depth of the circuit that prepares and measures the variational state $|\Psi\rangle$ is limited by noise. This effectively limits the size of fermionic systems that can be simulated on existing devices. To extend quantum simulations to subatomic systems beyond the deuteron~\cite{Dumitrescu2018} and Schwinger models beyond two spatial lattice sites~\cite{Martinez:2016yna,Klco:2018kyo}, we have recently proposed a preconditioning strategy~\cite{Klco:2018kyo} that transforms $\tilde{\mathcal{H}}_{SQ}$  into block-diagonal form by projecting it onto eigenstates of operators that represent good quantum numbers (e.g., parity, momentum, total spin) for the system of interest. As a result $\tilde{\mathcal{H}}_{SQ}=\bigoplus_{i} \mathcal{H}_{i}$, where $\mathcal{H}_{i}$ can now be interpreted as single-particle Hamiltonians acting on smaller subspaces than the original Hilbert space corresponding to $\tilde{\mathcal{H}}_{SQ}$. A Hamiltonian $\mathcal{H}_{i}$, specified by a $d\times d$ Hermitian matrix with elements $h_{kl}$ in some basis, can be mapped onto a Hamiltonian $\mathcal{H}^{i}_{QFP}$ that describes a frequency bin multiport device implementable with the QFP:
\begin{equation}\label{Eq:Ham}
\mathcal{H}^{i}_{QFP} = \sum\limits_{k=0}^{d-1} h_{kk} c^{\dagger}_{k}c_{k} + \sum\limits_{\substack{k,l=0 \\ k<l}}^{d-1} [h_{kl} c^{\dagger}_{k}c_{l} + h^{\ast}_{kl} c^{\dagger}_{l}c_{k}],
\end{equation}
where $h_{kl}$ are the entries in $\mathcal{H}_{i}$. In this encoding, we have mapped the original Hamiltonian $\tilde{\mathcal{H}}_{SQ}$ onto a set of single-particle systems defined by Eq.~(\ref{Eq:Ham}). To find ground-state energies of each single-particle Hamiltonian $\mathcal{H}_{i}$ we implement a variant of the VQE algorithm adapted for the QFP hardware. For the trial variational wavefunction $|\Psi\rangle$ we utilize an ansatz based on unitary coupled-cluster (UCC) theory
~\cite{mcclean2016}. The UCC wavefunction can be written as
\begin{equation}\label{Eq:UCC}
|\Psi\rangle = \exp\left(\sum\limits_{k=1}^{d-1}\theta_{k} [c^{\dagger}_{0}c_{k} - c^{\dagger}_{k}c_{0}]\right)|10\cdots0\rangle,
\end{equation}
where the state $\ket{0\cdots 1_k \cdots 0}$ denotes a single excitation (photon) in the frequency bin $k$, and none in the remaining $d-1$. The operator exponent can be evaluated explicitly in this case, leading to the following ($d{-}1$)-parameter state,
\begin{equation}\label{Eq:UCCstate}
|\Psi\rangle = \cos\phi\,|10\cdots0\rangle - \frac{\sin\phi}{\phi} \sum\limits_{k=1}^{d-1}\theta_{k}|0\cdots1_{k}\cdots0\rangle,
\end{equation}
with $\phi=\sqrt{\sum\limits_{k=1}^{d-1}\theta_{k}^{2}}$. In the context of the QFP, the UCC wavefunction $|\Psi\rangle$ represents a superposition of a single photon over $d$ frequency bins. 

With the Hamiltonian and UCC wavefunction defined, we use our QFP to estimate the expectation value $\langle \mathcal{H}^i_{QFP} \rangle = {\rm Tr} \left[|\Psi\rangle\langle\Psi|\mathcal{H}^i_{QFP}\right]$ for given parameter values $\{\theta_{k}\}$, by first preparing $|\Psi\rangle$ and experimentally reconstructing the elements of the single-particle density matrix $\rho_{kl} = (1/2)\langle\Psi|c^{\dagger}_{k}c_{l}+c^{\dagger}_{l}c_{k}|\Psi\rangle$. Measuring $\rho_{kl}$ is equivalent to placing the state $|\Psi\rangle$ on a 50/50 beamsplitter implemented between frequency bins $k$ and $l$, and recording the difference in the flux of detected particles in those modes immediately after the beamsplitter. Similarly, elements $\rho_{kk}$ can be measured by preparing the state $|\Psi\rangle$ and measuring the photon flux in each mode $k$ by using a photodetector. After repeating this process for all combinations of modes $k$ and $l$,  $\langle \mathcal{H}^i_{QFP} \rangle = {\rm Tr} [\rho \mathcal{H}^i_{QFP}] = \sum_{kl}\rho_{kl} h_{kl}$ can be estimated. Recent formulations of VQE, which use the current estimate of the energy $\langle \mathcal{H}^i_{QFP} \rangle$ to generate parameter updates $\{\delta\theta_{k}\}$ via a gradient-based classical optimizer, generally require many evaluations of $\langle \mathcal{H}^i_{QFP} \rangle$  to arrive at converged parameters.  We instead use a new method which merges the UCC ansatz with a many-body formalism called the anti-Hermitian contracted Schr{\"o}dinger equation\cite{mazziotti2006}.  This allows us to approximate the gradient of parameters using the measured $\rho_{kl}$ of each iteration~\cite{Morris2018} and arrive at convergence with significantly fewer evaluations of $\langle \mathcal{H}^i_{QFP} \rangle $. For example, in the problems we explore here our method required $\sim$20 iterations to converge compared to $\sim$500 iterations when using the BFGS algorithm~\cite{Fletcher1987}.

In practice, the measurement of the elements $\rho_{kl}$ using the single-photon state $|\Psi\rangle$ as an input is equivalent to a measurement with a coherent frequency comb where the relative amplitude of each comb line is set to $\theta_{k}\sin\phi/\phi$ (for lines $k=1,\ldots, d-1$) and $\cos\phi$ (for the line $k=0$) with respect to a reference coherent-state amplitude $\alpha$. Indeed, one can verify by a direct calculation that $\langle\Psi_\mathrm{comb}|\mathcal{H}^i_{QFP}|\Psi_\mathrm{comb}\rangle = |\alpha|^2 \langle\Psi|\mathcal{H}^i_{QFP}|\Psi\rangle$ where, 
\begin{equation}
|\Psi_\mathrm{comb}\rangle=\ket{\alpha\cos\phi}\otimes\Ket{\alpha e^{i\pi}\frac{\theta_{1}\sin\phi}{\phi}}\otimes\cdots\otimes\Ket{\alpha e^{i\pi}\frac{\theta_{d-1}\sin\phi}{\phi}}.
\end{equation}
Moreover, the measurements of $\rho_{kl}$ for non-overlapping sets of indices $k,l$ can be implemented in parallel, thus reducing the simulation time, as the QFP has an intrinsic ability to perform the same operation on different sets of modes in parallel. Previously, we implemented near-unity fidelity frequency-bin beamsplitters in parallel, with a theoretical predicted fidelity $\mathcal{F} = 0.9999$ and success probability $\mathcal{P} = 0.9760$~\cite{Lu2018a}. Such Hadamard gates can be achieved by driving two EOMs with $\pi$-phase-shifted sinewaves at frequency $\Delta\omega$ (with maximum temporal phase modulation $\Theta = 0.8169$ rad), and applying a step function with $\pi$-phase jump between the two computational modes on the central pulse shaper. The corresponding beamsplitter possesses $47.81\%$ reflectivity $\mathcal{R}$ (mode-hopping probability) and $49.79\%$ transmissivity $\mathcal{T}$ (probability of preserving frequency), with $2.4\%$ of the photons scattered outside of the computational space.

Despite such high fidelity, the residual imbalance in $\mathcal{R}$ and $\mathcal{T}$ is undesirable, leading to higher error in calculation of the $\rho_{kl}$ elements. Accordingly, in this work we further reduce the Hadamard gate's bias, achieving $\mathcal{R} = 48.7\%$ and $\mathcal{T} = 48.77\%$ (corresponding to a fidelity $\mathcal{F} = 0.999999$) by increasing $\Theta$ to $0.8283$ rad on both EOMs, while the QFP's central pulse shaper remains unchanged. The improved fidelity comes at the cost of greater photon leakage into adjacent sidebands ($2.53\%$ in this case) due to the stronger temporal modulation. Yet since possessing a beamsplitter with equal splitting ratio is more essential in this context than reducing leakage, we implement this higher $\mathcal{F}$, lower $\mathcal{P}$ Hadamard design. 

After setting up the Hadamard gates, we utilize the first pulse shaper to equalize the amplitude across all ten input frequency bins with the aid of the OSA. The relative spectral phase within every frequency pair is also fine-tuned until we find the in-phase condition as the reference---defined such that the lower (higher) frequency bin obtains the maximum (minimum) optical power after the Hadamard operation. To compute $\rho_{kl}$, we manipulate the relative amplitude and phase of a frequency pair $c_k$ and $c_l$, and record the optical power difference between two modes after the beam-splitting operation. To reconstruct the full density matrix, a total number of $d(d-1)/2$ beam-splitting operations is required in every iteration. Hence, the usage of five parallel beamsplitters (as well as the natural parallelization in pulse shapers and the OSA) reduces the required number of computations by a factor of five, before generating a new set of parameters $\{\theta^{\prime}_1,\cdots,\theta^{\prime}_M\}$ on the classical computer for the next iteration.

\paragraph*{Nuclear Structure Calculations}

Organizing principles that are rooted in the global symmetries of QCD have been successfully encoded in low-energy EFT frameworks describing nuclear forces, providing  a systematically improvable approach to calculations of atomic nuclei. At low resolution, i.e. at long 
wavelengths, details about the strong but short-ranged nuclear forces, or about QCD, are not revealed, and the lightest nuclei can be understood in 
terms of contact interactions of pairs and triplets of nucleons~\cite{bedaque2002,Kaplan:1998we,bedaque1999}.
In our model, we employ a Hamiltonian at next-to-leading order (NLO) in pionless EFT and adjust its parameters to the $S$-wave effective range expansions and the deuteron binding
energy; the strength of the three-body contact is adjusted to the triton binding energy. The Coulomb force  between protons is also included. We employ a finite basis consisting of eigenstates of the spherical harmonic oscillator with energy spacing $\hbar\omega=22$~MeV in a discrete variable representation~\cite{light2007,Bansal:2017pwn}, with the two-body and three-body potentials acting only between states with excitation energies up to and including 
 $2\hbar\omega$. This discretization maps nucleon fields $N$ onto annihilation operators $c_n$, with a corresponding interaction momentum cutoff of  337~MeV, leading to the second-quantized Hamiltonian $\mathcal{H}_{SQ}$. Next we project $\mathcal{H}_{SQ}$ onto Hilbert spaces with spin and parity of $J^{\pi}=0^{+}$ for $^4$He and  $J^{\pi}=1/2^{+}$ for $^3$H, and $^3$He, and the resulting Hamiltonian matrices $\mathcal{H}_{i}$ are implemented on
 the optical simulator, as described in the previous sections. 
\begin{figure}[!ht]
\centering
\includegraphics[width=0.32\textwidth]{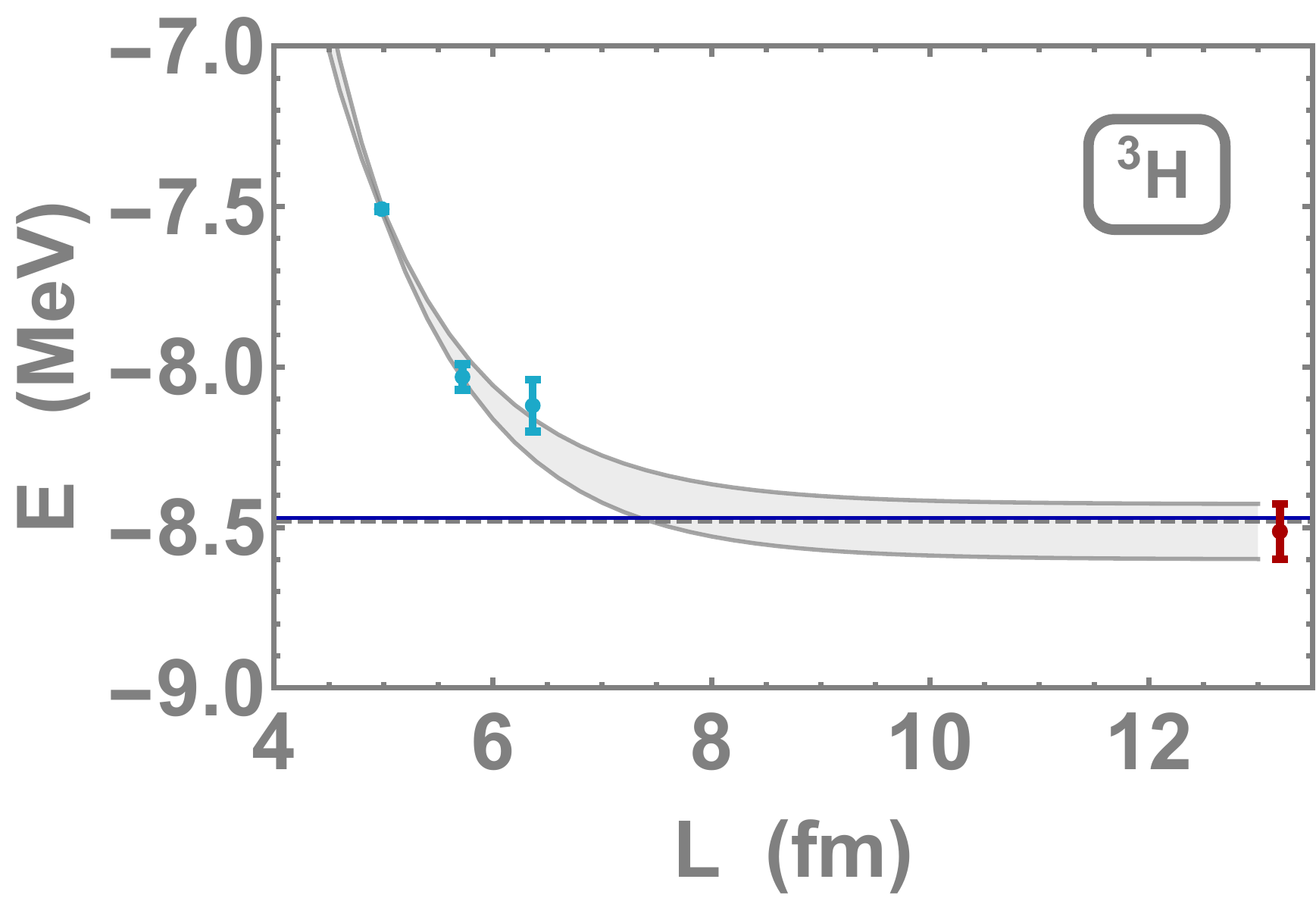}
\includegraphics[width=0.32\textwidth]{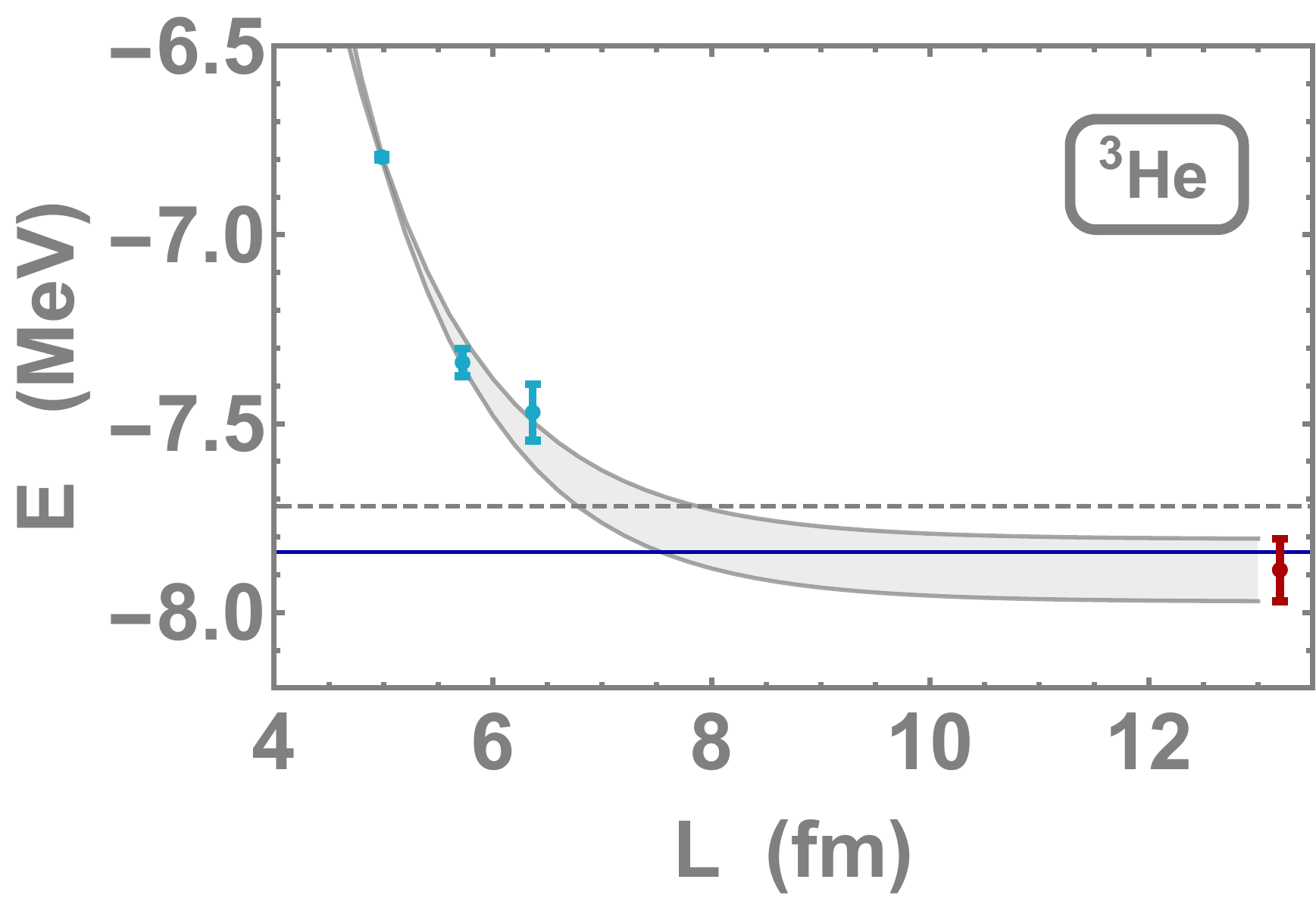}
\includegraphics[width=0.33\textwidth]{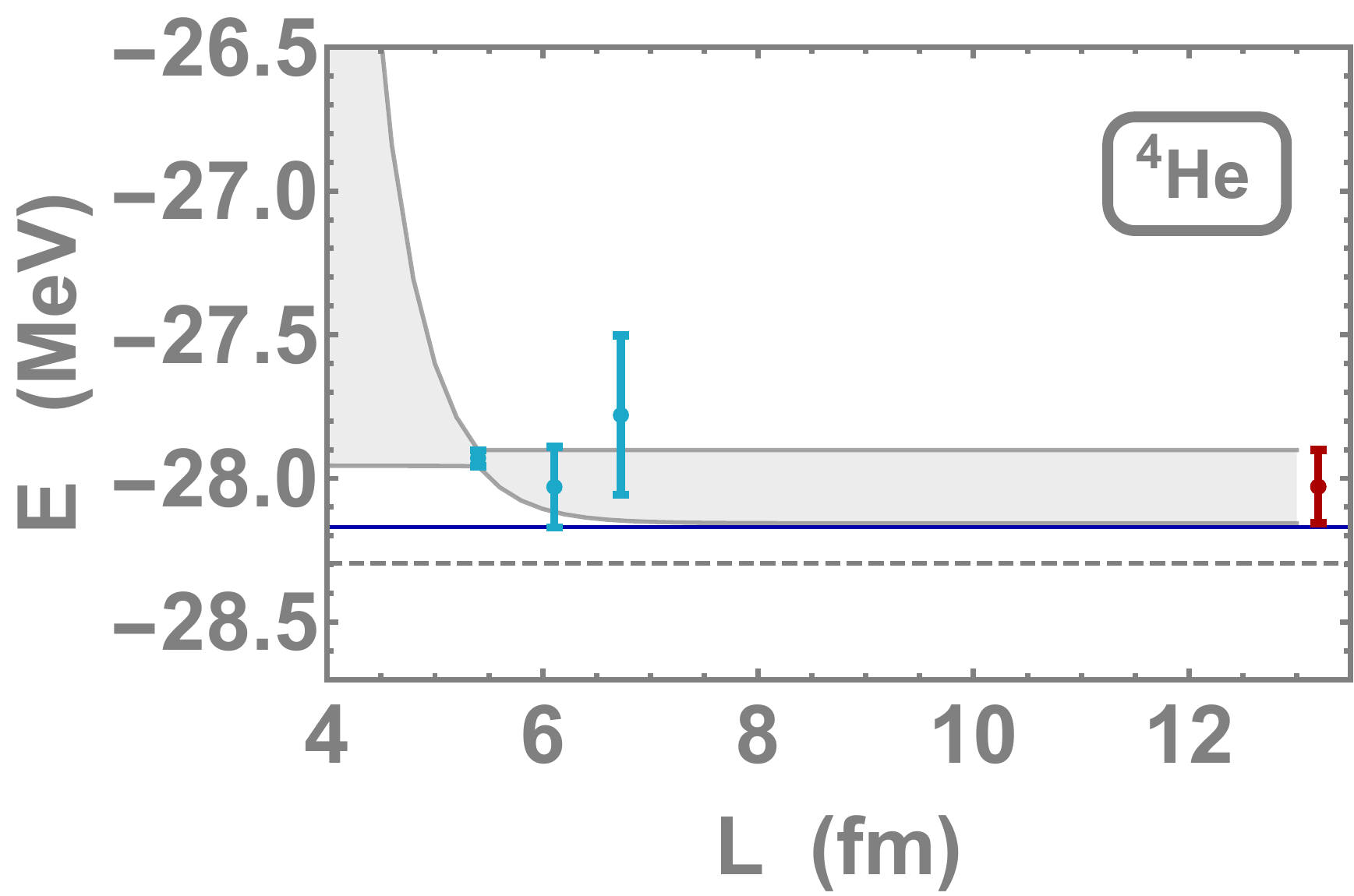}
\caption{
Results of the nuclear ground-state energies for $^3$H, $^3$He, and $^4$He nuclei computed with the QFP (blue data points) for Hilbert spaces with effective spatial extent $L$ with estimated systematic uncertainties. Also shown are the leading-order extrapolations via Eq.~(\ref{extra}) to infinite model spaces with propagated uncertainties (gray bands), the resulting extrapolated energies $E_\infty$ (red point at right), the tuned NLO EFT predictions
(dark blue solid line), and the known high-precision values of the binding energies (dashed gray line).
}
\label{fig:NSExtrap}
\end{figure}

Figure~\ref{fig:NSExtrap} shows the ground-state energies of $^3$H, $^3$He, and $^4$He computed with the VQE algorithm on the QFP, as a function of the effective spatial extent, $L$, of the model space.
For the weakly bound three-nucleon states of $^3$H and $^3$He, the energy is found to  decrease noticeably with increasing $L$. For these systems as well as $^4$He,  
extrapolations to large model spaces can be reliably performed
(shown as shaded regions)
using the leading-order expression given in Eq.~(\ref{extra}) of the Supplemental Material.
For each nucleus, the model-space-extrapolated binding energy 
(shown as the red point at right) is consistent with the corresponding tuned NLO EFT prediction. 
The uncertainties associated with each point, and in the extrapolated values, are dominated by systematic uncertainties in the QFP. We note that the underlying  Hamiltonian from the NLO pionless EFT is known to reproduce experimental data
to much better than the naive $\sim 10\%$ accuracy based upon power counting due to the relative size of coefficients in the effective range expansion.

\paragraph*{Schwinger Model Simulations}
Quantum Electrodynamics in 1+1 dimension, the Schwinger model~\cite{Schwinger:1962tp,Coleman:1975pw}, has been long-studied as an illuminating example of confinement and chiral symmetry breaking in quantum field theory, and is receiving new attention in the context of quantum computing and quantum information, for example Refs.~\cite{Banuls:2013jaa,Shimizu:2014fsa,Shimizu:2014uva,Banuls:2015sta,Banuls:2016lkq,Banuls:2016gid,Buyens:2016ecr,Buyens:2016hhu,Martinez:2016yna,Klco:2018kyo}. To represent this continuous theory on  computational devices, we employ the staggered discretization of the fermions~\cite{Kogut:1974ag} mapped to spin degrees of freedom using the Jordan-Wigner transformation. The resulting Hamiltonian $\tilde{\mathcal{H}}_{SQ}$ is detailed in Supplemental Material Eq.~(\ref{Eq:S3}). The electric field is  quantized and truncated between $\pm \Lambda$ (where the integer $\Lambda$ is chosen to ensure theoretical systematic uncertainties in the ground-state energy are below 1\%) and can create or annihilate positron-electron ($e^+e^-$) pairs while satisfying Gauss's law.
 Previous works have calculated the time evolution of 
 fluctuations in the electric field and charge density, 
 as well as static vacuum properties such as the chiral condensate on quantum devices~\cite{Martinez:2016yna,Klco:2018kyo}.  
 In this work, we extend the exploration to include  external static charges,
which are screened by 
deformations in the quantum vacuum. 
Being nondynamical, the static charges interact with the 
$e^+$'s and $e^-$'s 
 only through the electric field in their contribution to Gauss's law.  
Such systems are analogous to 
mesons found in nature containing a bottom or charm quark, 
and we denote these states as ``heavy mesons'' for convenience.
The Hilbert space of each vacuum sector is reduced by enforcing Gauss's law to contain only physical states and by projecting onto ground-state quantum numbers of 
parity and charge conjugation allowed by each static charge distribution.
The interaction potentials between static charges, and the effective ``nuclear'' forces between them, are extracted from  
combinations of ground-state energies. 
The associated modifications to the vacuum structure are isolated from differences in these ground-state wavefunctions.

The energy and wavefunction of the vacuum, of single static charges, of two like-sign and two-opposite sign static charges separated by a distance $r$, and of three static charges of the distinct charge combinations separated by distances $r_{12}$ and $r_{13}$ were calculated on the QFP by applying our VQE algorithm to Schwinger-model Hamiltonians $\mathcal{H}_{i}$ obtained by projecting the Hamiltonian $\tilde{\mathcal{H}}_{SQ}$  with eight fermion sites, corresponding to four spatial sites.
Using the ground-state energy solutions obtained on the QFP for all such Hamiltonians, we could then compute the heavy-meson mass, the potential energy between two and three static charges over a range of separations, 
and the fundamental two-body and three-body potentials. In addition, modifications to local observables due to the static charges could be determined.  
\begin{figure}
 \centering
\includegraphics[width=0.32\textwidth]{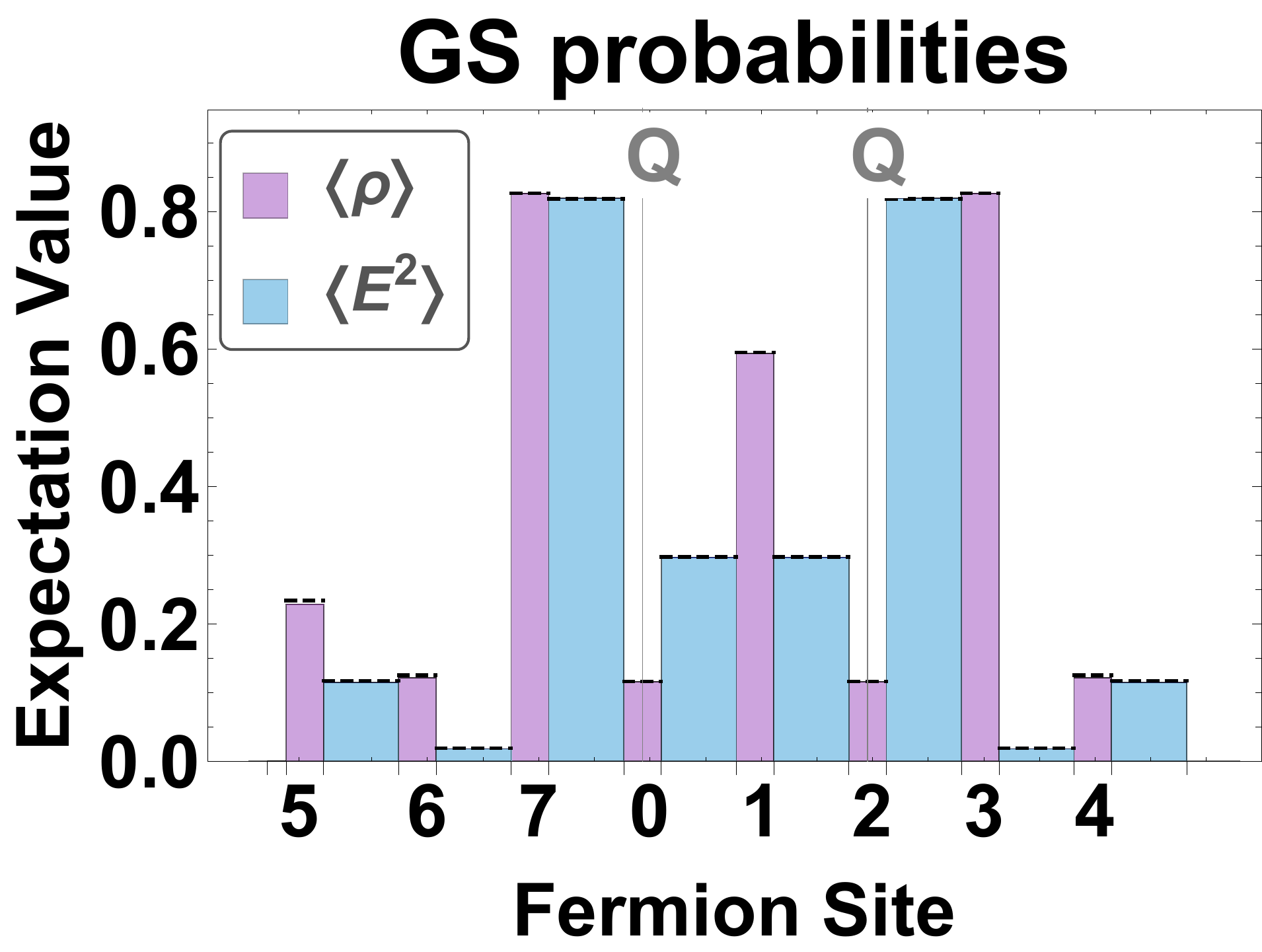}
\includegraphics[width=0.32\textwidth]{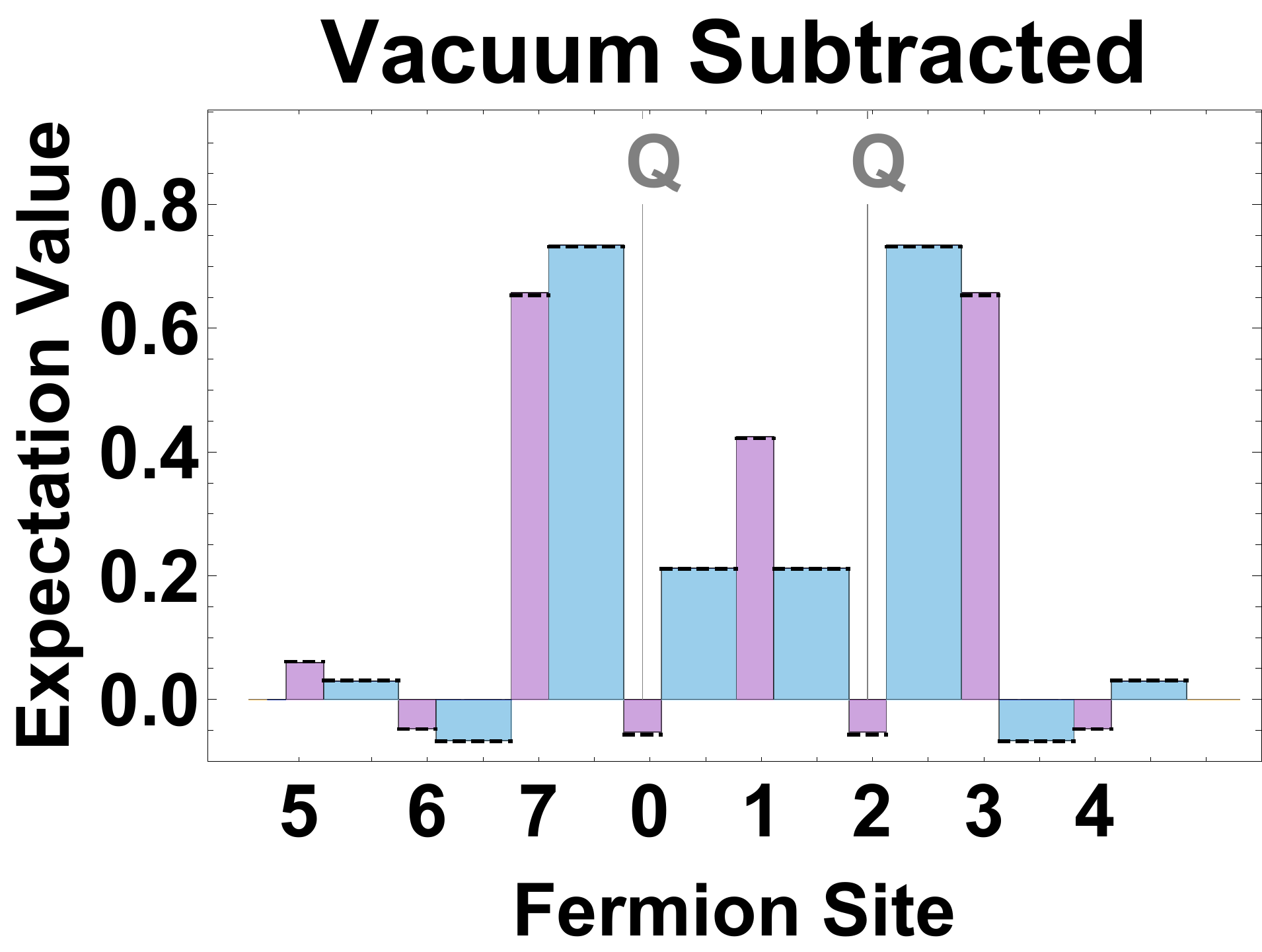}
\includegraphics[width=0.32\textwidth]{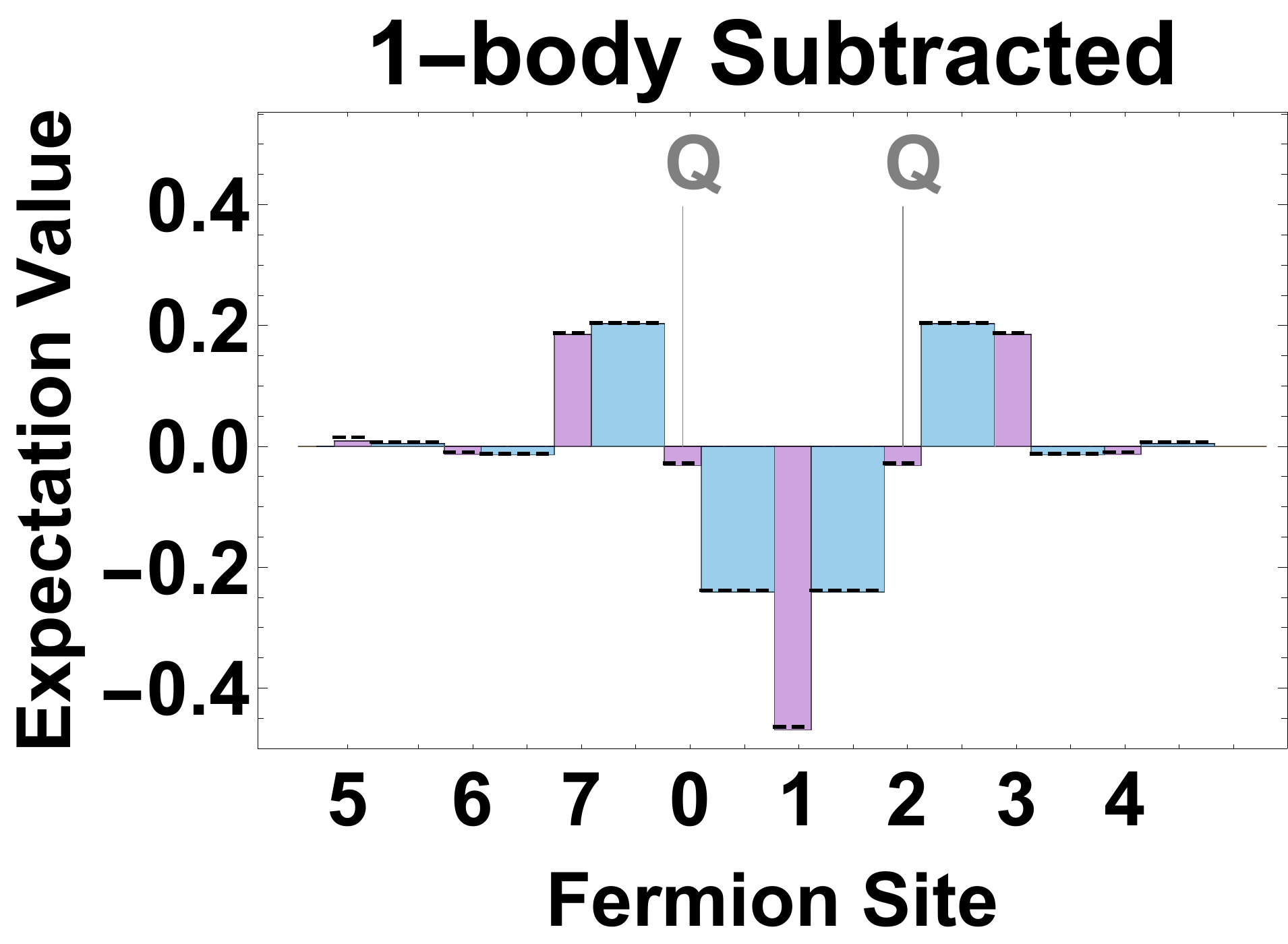}
\caption{
The $e^+e^-$ densities ($\langle\rho\rangle$) and energy-density in the electric field ($\langle E^{2}\rangle$) for $Q=-1$ static charges separated by one spatial lattice site.
The left panel shows the raw distributions, the center panel has the vacuum removed, and the right panel also has the contributions from the individual charges removed.
The horizontal black dashed lines are the analytic values of the local densities, while the error bands (not seen on this scale) represent fluctuations over the last ten iterations of the VQE.  The values shown in each panel are presented in Table~\ref{tab:localr2} in the Supplemental Material.
}
\label{fig:VBBr2dist}
\end{figure}
As an example, Fig.~\ref{fig:VBBr2dist} shows the $e^+$ and $e^-$ probabilities per spatial site and the energy density of the electric field for the ground state of two static negative charges separated by one site, computed from the VQE results. Localized modifications to the vacuum resulting from the static charges are clearly discernible. Similar sets of plots can be generated from the VQE solutions for all other two- and three-charge configurations 
(see Supplemental Material). 
From these, the full potential energy can be computed.
\begin{figure}[!ht]
\centering
\includegraphics[width=0.97\textwidth]{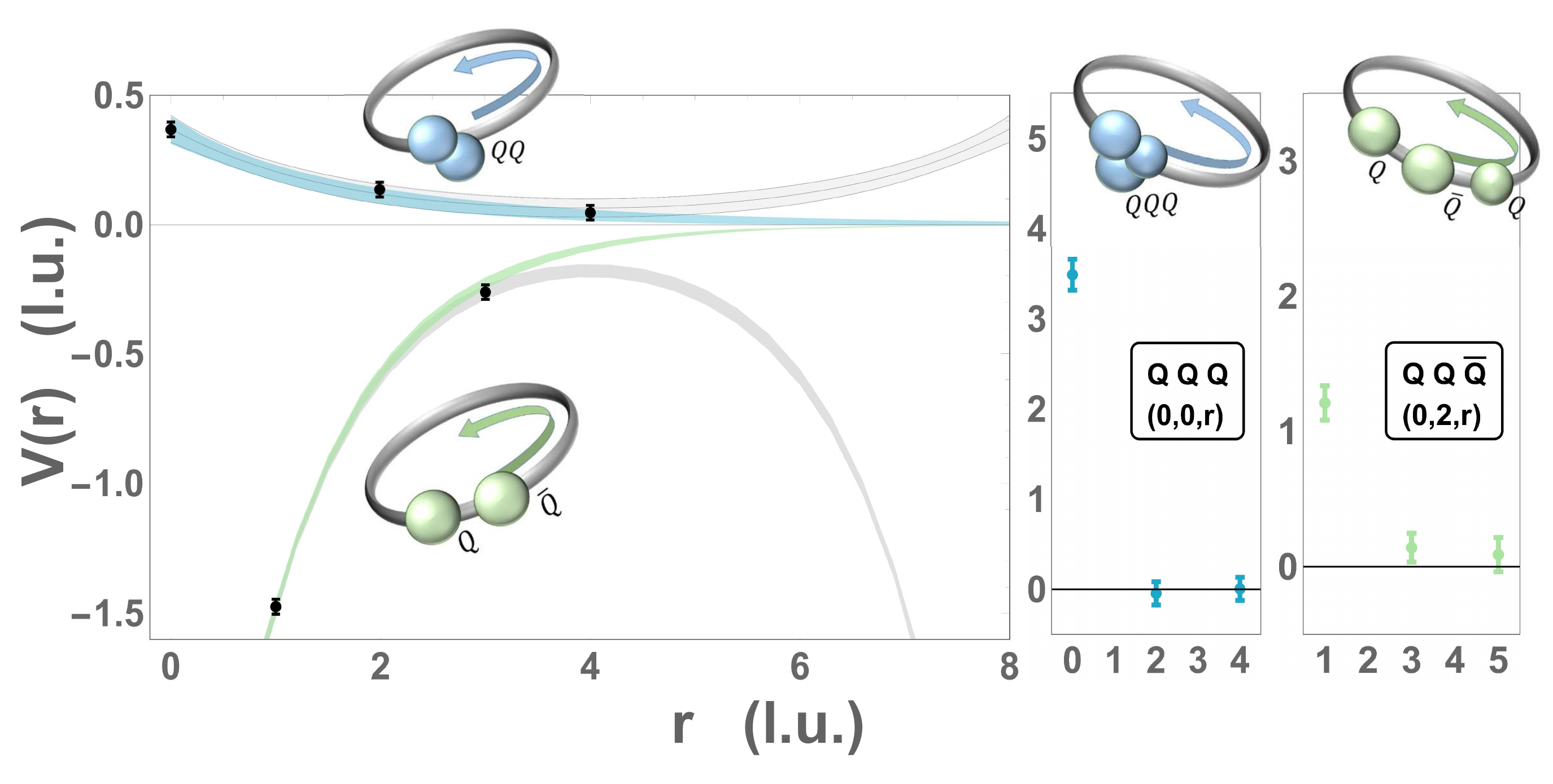}
\caption{
The left panel shows 
the potential between two like-sign static charged sources (upper curves) and between opposite-sign charges (lower curves) as a function of 
separation (in lattice units). 
The symmetric gray curves represent the extracted lattice potential including the presence of image charges.  
The blue and green bands represent the infinite-volume potentials using correlated extracted values of the couplings and masses 
(see Supplemental Material
for details).
The center panel shows the three-body potential for 
three like-sign static charges.  The right panel shows the three-body potential for two like-sign and one opposite-sign static charges.
}
\label{fig:VrBBexpt}
\end{figure}
Figure~\ref{fig:VrBBexpt} (left) shows the potential energy as a function of separation
between two  like-sign and opposite-sign static charges. (The systems in Fig.~\ref{fig:VBBr2dist} contribute specifically to the point at $r=2$.)  For distances that are large compared with the radius of the heavy meson, the potential between the static charges is expected to fall as the sum of exponentials with arguments set by the light-hadron spectrum.  
Fitting to the simulation results from the QFP, including the effect of image charges from the boundary conditions, and isolating the infinite-volume limit, gives the phenomenological fit potentials in Fig.~\ref{fig:VrBBexpt}. The short-distance forms of the potentials are expected to deviate from these forms, but such deviations are expected to 
be a small effect in subsequent analyses.
These phenomenological potentials are used in the Schr\"odinger equation to match to a low-energy EFT description 
of the Schwinger model 
in terms of local contact operators, the analogue of the pionless theory,
${\cal H}=-C_N (N^\dagger N)^2+ ...$.
The EFT parameters can be determined for a given  heavy-meson mass, but
higher-order interactions cannot be meaningfully 
constrained with the present calculations.
The three-body potentials between 
static charges 
 are extracted in similar ways to the two-body potentials, with the two-body potentials removed. Two slices of the retrieved three-body potentials are displayed in
Fig.~\ref{fig:VrBBexpt} (right two panels). 
As expected, these potentials fall rapidly as  the bodies  separate.  Phenomenological three-body potentials can be extracted and used to constrain the coefficients of three-body EFT parameters.

The present calculations have been performed with one lattice spacing and one spatial volume, and 
calculations in systems with more lattice sites and smaller lattice spacing would provide higher resolution of the extracted potentials and more precise values of the
corresponding EFT parameters.

\paragraph*{Discussion}

Establishing a direct connection between the fundamental building blocks of our universe, the quarks and gluons, and the properties and dynamics of matter under a range of conditions, from those created in the laboratory to extreme astrophysical environments, faces challenges beyond the capabilities of foreseeable classical computation.
From exponentially-growing Hilbert spaces required to describe nuclei, to sign problems in evaluating finite density systems, anticipated developments in 
quantum devices and quantum information offer the hope of addressing these Grand Challenge problems in subatomic physics. 

VQE algorithms implemented with error-corrected qubits are anticipated to provide a scalable path to solving these Grand Challenge problems on quantum devices in the future. However, currently-available hardware is too noisy to demonstrate such quantum advantages. In this work, we explored a way to implement VQE optically by using the QFP, with classical computers to reduce computational complexity of quantum simulations before QFP implementation. In particular, we use classical resources to project many-body fermionic Hamiltonians corresponding to nuclear and quantum field theory systems onto a hierarchy of single-particle Hamiltonians that can be simulated
efficiently on the QFP. While the single-particle (photon) calculations on the QFP utilize states with  quantum correlations, the required resources scale similarly to those of classical simulations.  This demonstration of controlled, single-photon-equivalent manipulation is a first step towards scalable QFP simulations where input states are modified to consist of multiple photons.  Together, the QFP and such state preparation of higher complexity is expected to require resources that scale polynomially with the size of the quantum system and thus exhibit a quantum advantage.

In this work, we have presented results 
from the largest photonics-based quantum simulation, using an  all-optical  quantum  frequency processor, to demonstrate the potential of this technology for calculations in subatomic physics.  
We presented the two-body and three-body interactions between composite objects,
and resulting terms in its low-energy EFT,
in the Schwinger model, which shares characteristics with QCD.
Importantly, representing a key ingredient in the connection between quarks and gluons and nuclei,  
a low-energy EFT of QCD was used to calculate the binding energies of $^3$H, $^3$He,  and $^4$He.
While the results of our calculations are not of comparable complexity or precision to those that can be achieved today with classical computation, they are an encouraging first step in exploring the utility of hybrid  optical quantum devices for addressing Grand Challenges in subatomic physics.

\bibliography{master}

\section*{Acknowledgments}
NK and MJS would like to thank Silas Beane, David Kaplan and Aidan Murran for valuable discussions.
This material is based upon work supported by the U.S. Department of
Energy, Office of Science, Office of Nuclear Physics under Award
Nos. DEFG02-96ER40963 (University of Tennessee), DE-SC0018223 (SciDAC-4 NUCLEI). 
NK and MJS were supported  by DOE grant No.~DE-FG02-00ER41132. NK was supported in part by the Seattle Chapter of the Achievement Rewards for College Scientists
(ARCS) foundation. This work is supported by the U.S. Department of Energy, Office of Science, Office of Advanced Scientific Computing Research (ASCR) quantum algorithm teams and testbed programs, under field work proposal numbers ERKJ333 and ERKJ335. AE received funding from the European Research Council (ERC) under the European Union's Horizon 2020 research and innovation programme (grant agreement No 758027). This work was performed in part at Oak Ridge National Laboratory, operated by UT-Battelle for the U.S. Department of Energy under Contract No. DEAC05-00OR22725.

\newpage
\section*{Supplementary materials}
Supplementary Text\\
Figs. S1 to S5\\
Tables S1 to S9\\
References \textit{(71-87)}

\setcounter{equation}{0}
\setcounter{figure}{0}
\setcounter{table}{0}
\setcounter{page}{1}
\renewcommand{\theequation}{S\arabic{equation}}
\renewcommand{\thetable}{S\arabic{table}}
\renewcommand{\thefigure}{S\arabic{figure}}

\paragraph*{Results}
\FloatBarrier
The following section provides details of the VQE implementation on the QFP for the light nuclei and Schwinger model systems presented in the main text.
\begin{table}[H]
  \centering
  \begin{tabular}{ccccc}
  \hline
  \hline
     & $d$ & Comp./Iter. & VQE energy & Time (min.) \\
     \hline
     $^3$H ($N_\mathrm{max} = 2$) & 5 & 10 & $-$7.5075(1) & 2.3   \\
     $^3$H ($N_\mathrm{max} = 4$) & 15 & 105 & $-$8.031(6) & 27.1 \\
     $^3$H ($N_\mathrm{max} = 6$) & 34 & 561 & $-$8.12(2) & 134.6 \\
     $^3$He ($N_\mathrm{max} = 2$) & 5 & 10 & $-$6.7942(1) & 2.3 \\
     $^3$He ($N_\mathrm{max} = 4$) & 15 & 105 & $-$7.3380(3) & 26.9 \\
     $^3$He ($N_\mathrm{max} = 6$) & 34 & 561 & $-$7.470(9) & 133.8 \\
     $^4$He ($N_\mathrm{max} = 2$) & 5 & 10 & $-$27.9301(2) & 2.2 \\
     $^4$He ($N_\mathrm{max} = 4$) & 20 & 190 & $-$28.03(1) & 48.6 \\
     $^4$He ($N_\mathrm{max} = 6$) & 64 & 2016 & $-$27.78(2) & 500\\
   \hline
  \hline
  \end{tabular}
  \caption{
  Simulation results for the ground-state energies of the light nuclei at NLO in the pionless EFT obtained with the QFP.  The first two columns designate the nucleus (model space) and the dimensionality of the Hilbert space.  The third column indicates the computations per iteration. The energy and associated standard deviation are quoted from the last five iterations of a converged VQE (statistical uncertainty only).  The final column indicates the total time to complete 25 iterations.}
\end{table}

\begin{table}[!ht]
  \centering
  \begin{tabular}{ccccc}
  \hline
  \hline
     & $d_\mathrm{sym}$ & Comp./Iter. &  VQE energy & Time (min.) \\
     \hline
     Vac &  9 & 15 & $-2.01503(1) $& 7.1 \\
     (0) &  26 & 154 & $-0.73167(2) $& 75 \\
     (0,0) &  9  & 21 & $0.91843(2) $& 11.4 \\
     (0,2) &  16 & 42 & $0.68691(1) $& 20.1 \\
     (0,4)$^{\dagger}$ &  17 & 45 & $0.59856(6) $& 67.0 \\
     (0,1) &  41 & 225 & $-0.92433(2) $& 99.3 \\
     (0,3)$^{*}$ &  58 & 222 & $0.289190(9) $& 298 \\
     (0,0,0) & 7 & 15 & $6.42752(8) $& 7.3 \\
     (0,0,2) & 5 & 9 & $2.428332(1) $& 4.5 \\
     (0,0,4) & 5 & 9 & $2.300160(2) $& 4.7 \\
     (0,2,4) & 5 & 9 & $2.210773(3) $& 4.7 \\
     (0,0,1) & 59 & 385 & $0.5418(4) $& 173 \\
     (0,0,3) & 31 & 87 & $1.83890(1) $& 40.1 \\
     (0,2,1) & 35 & 217 & $0.22861(4) $& 99.6 \\
     (0,2,3) & 62 & 406 & $0.3731(1) $& 194.6 \\
     (0,2,5) & 24 & 105 & $1.535129(9) $& 46.9 \\
     (0,4,1) & 68 & 448 & $0.3466(3) $& 196.8 \\
  \hline
  \hline
  \end{tabular}
  \caption{Simulation results for ground-state energies in the  Schwinger model obtained using the QFP.  The first two columns designate the configuration of static charges and the symmetry-projected dimensionality of the Hilbert spaces.  The third column indicates the computations per iteration [less than $d_\mathrm{sym}(d_\mathrm{sym}-1)/2$ due to matrix sparsity].  The energy and associated standard deviation are quoted from the last ten iterations of a converged VQE.  The final column indicates the total time to complete 50 iterations. ($^{\dagger}$) Obtained with three beamsplitters and 100 iterations. ($^{*}$) Obtained with 150 iterations.}
\end{table}

\FloatBarrier
\paragraph*{Nuclear Structure}

A  Hamiltonian from the NLO pionless EFT~\cite{vanKolck:1998bw,Kaplan:1998we,Kaplan:1998tg,bedaque2002}, a systematically-improvable approach to nuclear interactions at low energies,  is employed for computations of the nuclei $^3$H, $^3$He,  and  $^{4}$He, with a  momentum-space potential
\begin{eqnarray}\label{Eq:S1}
V &=& V^{({}^1S_0)}_{NN}(p',p) + V^{({}^3S_1)}_{NN}(p',p) + V_{NNN} \ , \nonumber\\ 
V_{NN}^{({}^1S_0)}(p',p) &=& \tilde{C}_{^1S_0} + C_{{}^1S_0}({p'}^2  +p^2) \ , \nonumber\\
V_{NN}^{({}^3S_1)}(p',p) &=& \tilde{C}_{^3S_1} + C_{{}^3S_1}({p'}^2  +p^2) \ , \nonumber\\
V_{NNN} &=& \frac{c_E}{F_\pi^4\Lambda_\chi}\sum_{1\le i\ne j\le A} \vec{\tau}_i\cdot\vec{\tau}_j \ .
\end{eqnarray}
%
Here, $p$ and $p'$ denote magnitudes of the incoming and outgoing relative three-momentum, respectively. The nucleon-nucleon  potentials $V_{NN}$ act in the singlet and triplet $S$-waves with $\tilde{C}_{^1S_0}=-0.7617$~MeV$^{-2}$, $C_{^1S_0}=2.9098$~MeV$^{-4}$,  $\tilde{C}_{^3S_1}= -1.2014$~MeV$^{-2}$, and  $C_{^3S_1}=3.3984$~MeV$^{-4}$, respectively. 
These parameters were determined by fits to the effective range expansion in the respective partial waves and to the deuteron binding energy.
The three-nucleon potential $V_{NNN}$ employs the isospin operators $\vec{\tau}_i$ for the nucleon $i$, the parameters $F_\pi=92.4$~MeV and $\Lambda_\chi=700$~MeV. The parameter  $c_E=0.01929$ is adjusted to reproduce the triton binding energy.   
This EFT is implemented as a discrete variable 
representation (DVR)~\cite{light1985,baye1986,littlejohn2002} in the harmonic oscillator basis, using translationally-invariant Jacobi coordinates and the infrared corrections of Ref.~\cite{binder2016,Bansal:2017pwn}. The potentials act only 
between states with excitation energies up to and including 2$\hbar\omega$, while the kinetic energy is not truncated. 
While our results are insensitive to the ultraviolet cutoff of the potential, we have chosen to work with a cutoff of  
337~MeV and $\hbar\omega=22$~MeV.
\begin{figure}[!ht]
    \centering
    \includegraphics[width=0.45\textwidth]{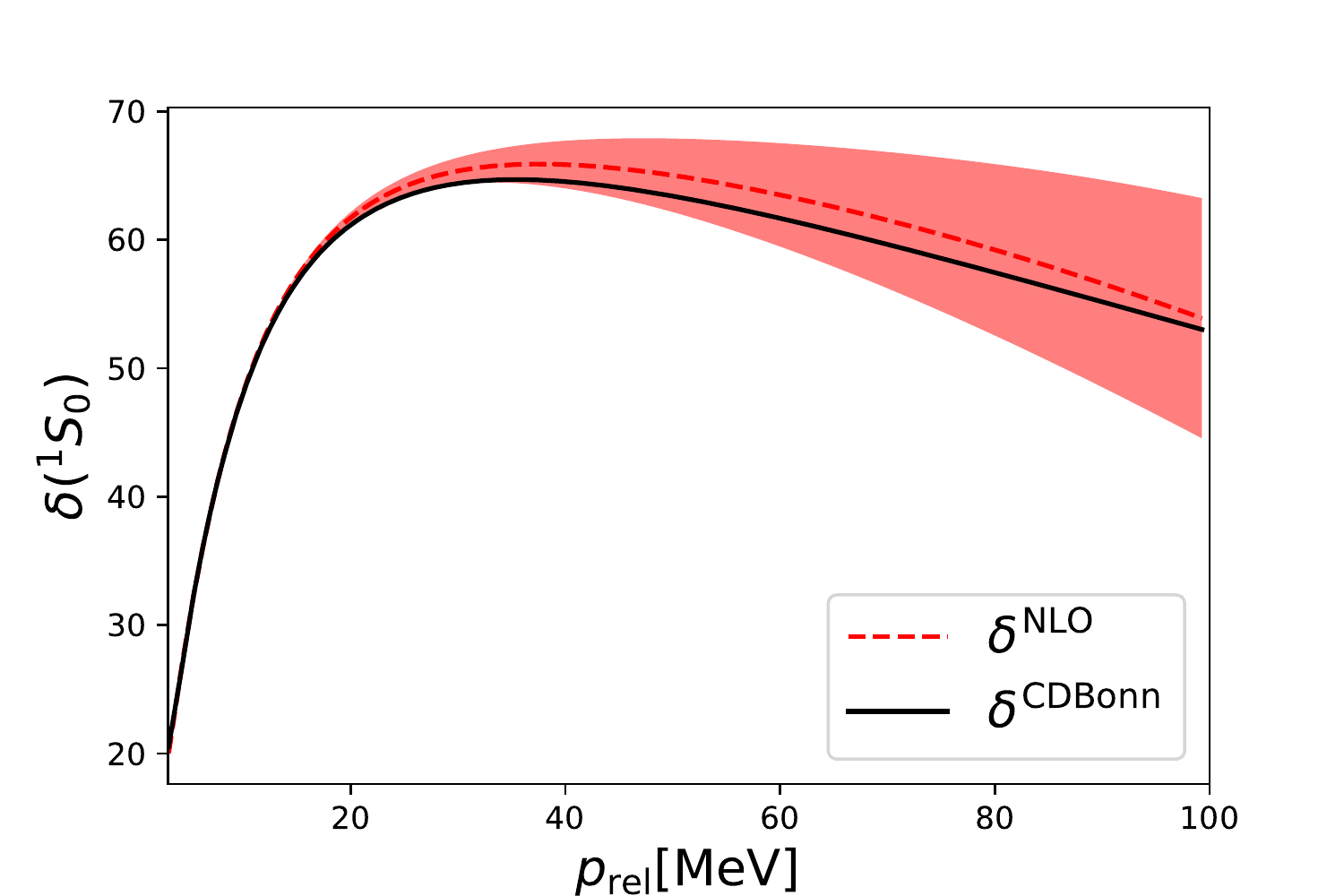}
    \includegraphics[width=0.45\textwidth]{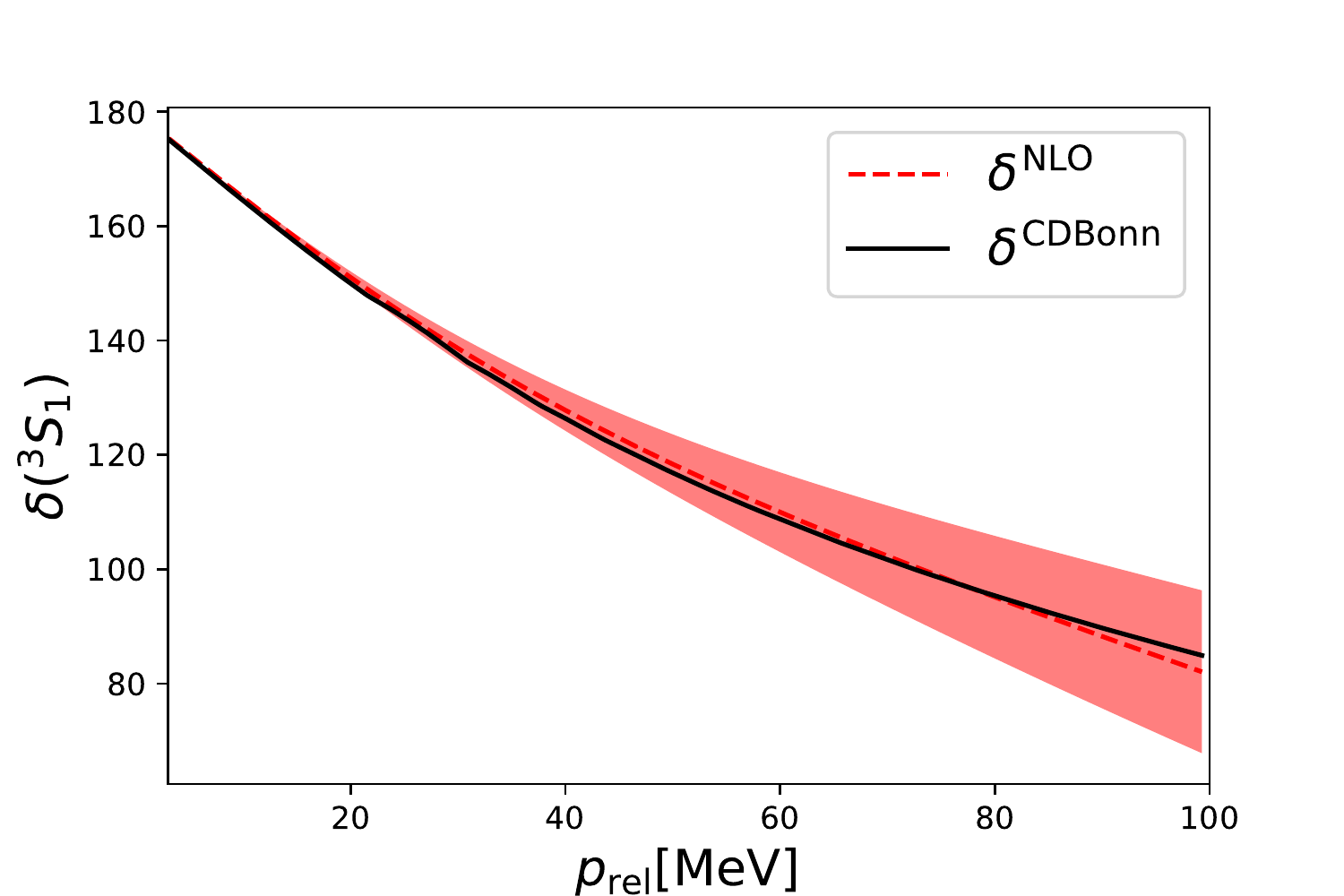}
    \caption{
    Nucleon-nucleon phase shifts in the singlet (left) and triplet (right) $S$-wave channels (in degrees) as a function of relative momentum.  
    The solid (black) lines correspond to high-precision results obtained with the CD Bonn potential.
    The dashed (red) lines correspond to results obtained from the NLO pionless EFT, while the shaded (red) region includes estimates from higher orders in the pionless EFT expansion that are not included in the interactions employed on the QFP.
    }
    \label{fig:phaseshifts}
\end{figure}
Figure~\ref{fig:phaseshifts} shows phase shifts in the singlet and triplet $S$-wave channels
obtained from the pionless theory with matched EFT parameters, and compares them to the corresponding phase shifts obtained from the high-precision CD Bonn potential~\cite{machleidt2001}. 
Our NLO EFT potentials 
reproduce the phase shifts within uncertainty estimates.
These systematic theory uncertainties, shown as shaded regions in Fig.~\ref{fig:phaseshifts}, result from  
estimates of the contributions from terms 
that are higher-order (NNLO) in the pionless EFT but are not included in our calculations.

Using standard tools~\cite{navratil2000b,navratil2007},
Hamiltonian matrices in Hilbert spaces 
with spin and parity $J^\pi=1/2^+$ and $0^+$ for the three-nucleon systems $^3$H, $^3$He, and $^4$He, respectively, are computed. 
Limiting the total number of
harmonic oscillator quanta to
$N_{\rm max}=2,4,6$ 
leads to 
 dense Hamiltonian matrices of dimension $d=5, 15, 34$ and $d=5, 20, 64$ for the
nuclei with mass numbers $A=3$ and $A=4$, respectively. These Hamiltonian matrices provide the inputs for the VQE algorithm.

At low energies, states projected onto a finite harmonic oscillator basis closely resemble those projected onto modes of a spherical cavity with a hard-wall radius $L\sim [N_{\rm max}/(m\omega)]^{1/2}$~\cite{furnstahl2012}. It can be shown~\cite{furnstahl2012}, using a development that parallels that of L\"uscher and others to establish finite-volume corrections to localized states in lattice QCD calculations~\cite{Luscher:1985dn,Luscher:1986pf},
that the leading finite-model-space shifts in the energy of an isolated bound state have the form,
\begin{equation}
E(L)=E_\infty +a  e^{-2k_\infty L}   
 .
\label{extra}
\end{equation}
From bound state energies $E(L)$ determined in three finite model spaces, and the separation momentum 
\begin{equation}
k_\infty = \frac{1}{\hbar}\sqrt{-2m[E_\infty(A) -E_\infty(A-1)]}    
\label{eq:kErel}
\end{equation}
of the $A$-body system~\cite{konig2018,forssen2018}, we determine the amplitude $a$ and the infinite-model-space energy $E_\infty$ by a fit.
Using the values of $L$ tabulated in the Supplementary Material of Ref.~\cite{wendt2015}, 
the binding energies $E(L)$ computed in finite model spaces using our QFP,
that are shown in Table~\ref{tab:nuclei},
are extrapolated to infinite model spaces.

\begin{table}[!ht]
  \centering
  \begin{tabular}{l|lll|lll}
  \hline
  \hline
                & \multicolumn{3}{c|}{Quantum frequency processor} & \multicolumn{3}{c}{Exact diagonalization}\\
  $N_{\rm max}$ & $^3$H    & $^3$He   & $^4$He    & $^3$H    & $^3$He   & $^4$He\\
     \hline
     2          & $-7.508(8)$ & $-6.794(7)$ & $-27.93(3)$ & $-7.513$ & $-6.800$ & $-27.947$ \\
     4          & $-8.031(40)$ & $-7.338(37)$ & $-28.03(14)$ & $-8.060$ & $-7.366$ & $-28.106$  \\
     6          & $-8.120(81)$  & $-7.470(75)$ & $-27.78(28)$ & $-8.275$ & $-7.600$ & $-28.148$  \\
     $N_A$      & ---      & ---      &  ---     & $-8.482$ & $-7.830$ & $-28.165$ \\
     \hline
     $\infty$   & $-8.51(9)$  & $-7.89(8)$  & $-28.04(14)$ & $-8.47$  & $-7.84$  & $-28.17$      \\
     \hline
     Exp.       & $-8.482$ & $-7.718$ & $-28.296$ & $-8.482$ & $-7.718$ & $-28.296$ \\ 
  \hline
  \hline
  \end{tabular}
  \caption{
  Ground-state energies of light nuclei obtained from Hamiltonian diagonalization of the NLO pionless EFT using the QFP compared to the exact results for model-spaces of size $N_{\rm max}=2,4,6$, and their extrapolation to infinite model space, $N_{\rm max}=\infty$. For comparison, the experimentally known values are  given (Exp.), along with the results obtained from exact diagonalizations in a  large model space, $N_A$.  For the quantum frequency processor, systematic simulation uncertainties of (0.1\%, 0.5\%, 1\%) in the VQE for $N_{\rm max}$ of (2, 4, 6) are extrapolated to the infinite model space  as shown in Fig.~\ref{fig:NSExtrap} of the main text.  This extrapolation uses the form of Eq.~(\ref{extra}) defining $k_\infty$ through the separation energy and enforcing the constraint $a<1$ GeV.
  }
  \label{tab:nuclei}
\end{table}
Table~\ref{tab:nuclei} also shows the results of exact diagonalizations of the Hamiltonian matrices describing the light nuclei in Hilbert spaces with a range of $N_{\rm max}$, as indicated. Diagonalizations in large model spaces of size $N_A$ 
($N_3=40$ and $N_4=20$ for nuclei with mass number $A=3$ and $A=4$, respectively) are essentially converged with respect to the size of the model space. The infinite-model-space extrapolation results obtained by fitting Eq.~(\ref{extra}) to the data obtained
with $N_{\rm max}=2,4,6$ are also shown;
these results are close to the (essentially) exact numerical results, and higher-order corrections to the leading-order result~(\ref{extra}) are suppressed by powers and exponentials of
$k_\infty L$~\cite{furnstahl2014}. 
Our extrapolated results for $^{3}$He and $^4$He are close to their experiment values, consistent with expectations from a leading-order Hamiltonian within statistical and systematic uncertainties. 
(We remind the reader that the three-body EFT parameter was adjusted to reproduce the ground-state energy of $^3$H.) 
The results obtained from the QFP, that are shown in Fig. 3 of the main text, are also given in Table~\ref{tab:nuclei}.
Through repeated measurements using the QFP, we have identified a systematic uncertainty of $\pm 1\%$ that accompanies each measurement, which is significantly larger than the associated statistical uncertainties.
In extrapolating to infinite model space, this systematic uncertainty of each point is uniformly sampled over in performing a Monte Carlo to determine the uncertainty of each binding energy as a function of $L$, including the $L\rightarrow\infty$ limit.  
For each nucleus, a value of the binding for each of the three measured $L$ values is uniformly sampled from the interval arising from the systematic uncertainty.  These samples are then used in a two-parameter fit of the form in Eq.~(\ref{extra}), where the difference in binding energies are used to relate $k_\infty$ to $E_\infty$ via Eq.~(\ref{eq:kErel}). These fit parameters define a curve as a function of $L$.  This process is repeated a large number of times to establish the shaded fit region and infinite-model-space value.

\paragraph*{Lattice Hamiltonians for Schwinger Model simulations}

In 1+1 dimensional quantum electrodynamics, latticized with staggered fermions~\cite{Kogut:1974ag}, and transformed to spin degrees of freedom using the Jordan-Wigner transformation, the Hamiltonian of the Schwinger model~\cite{Schwinger:1962tp,Coleman:1975pw} can be written as
\begin{equation}\label{Eq:S3}
\tilde{\mathcal{H}}_{SQ} = x \sum_{n = 0}^{N_{fs} -1} \left( \sigma_n^+ L_n^- \sigma_{n+1}^- + \sigma_{n+1}^+ L_n^+ \sigma_n^- \right) + \sum_{n = 0}^{N_{fs} -1} \left( \ell_n^2 + \frac{\mu}{2} \left(-\right)^n \sigma_n^z \right) \ \ \ .
\end{equation}
The kinetic (hopping) term contains raising and lowering operators, $L^\pm$, modifying the value of the electric field that is naturally quantized 
(between truncations $\pm \Lambda$) in one dimension. Choosing periodic boundary conditions for this one-dimensional spatial lattice produces a Hamiltonian with discrete, rotational symmetry.  While this representation is perfectly suited for qubit implementation---creating a latticized, tensor-product structure with single qubits at the sites to represent fermion occupation and registers of $\lceil\log_2(2\Lambda+1)\rceil$ qubits on each link for the electric field---the additional constraint of Gauss's law makes this representation both excessive for physical states and sensitive to noise within the quantum computation.  
Instead,  the lattice configurations in the physical sector (that satisfy Gauss's law) are classically enumerated and mapped onto quantum states of the Hamiltonian.  Because of the locality of interactions, the Hamiltonian remains sparse in this representation. By working only with configurations in the physical subspace, the Hilbert space dimension in terms of $N_s$, the number of spatial sites, is reduced from $e^{\log(64) N_s}$ to $1.02(1)e^{1.1772(2) N_s}$ and the four-spatial-site lattice becomes  accessible to our QFP.
\begin{table}[!ht]
  \centering
  \begin{tabular}{cccccc}
  \hline
  \hline
     & $\Lambda$ & $d$ & Symmetries & $d_\mathrm{sym}$ & $E_{GS}$ precision \\
     \hline
     Vac & 3 & 53 & P, $\vec{p}$ & 9 & 0.2\% \\
     (0) & 4 & 50 & P & 26 & 0.4\% \\
     (0,0) & 5 & 16 & P & 9 & 0.2\%\\
     (0,2) & 5 & 31 & P & 16 & 0.09\% \\
     (0,4) & 5 & 35 & P & 17 & 0.05\% \\
     (0,1) & 5 & 67 & CP & 41 & 0.8\% \\
     (0,3) & 8 & 95 & CP & 58 & 0.15\% \\
     (0,0,0) & 12 & 13 & P & 7 & $5\times 10^{-7}$\% \\
     (0,0,2) & 5 & 5 & & 5 & 0.04\% \\ 
     (0,0,4) & 4 & 5 & & 5 & 0.6\% \\
     (0,2,4) & 4 & 9 & P & 5 & 0.1\% \\
     (0,0,1) & 7 & 59 & & 59 & 0.13\% \\
     (0,0,3) & 7 & 31 & & 31 & 0.7\% \\
     (0,2,1) & 6 & 67 & P & 35 & 0.6\% \\
     (0,2,3) & 7 & 62 & & 62 & 0.1\% \\
     (0,2,5) & 7 & 46 & P & 24 & 0.74\% \\
     (0,4,1) & 7 & 68 & & 68 & 0.1\% \\
  \hline
  \hline
  \end{tabular}
  \caption{
  Properties of the systems studied with our quantum device.  The first column indicates the locations of static charges [with charge $-Q$ ($+Q$) for odd (even) sites, respectively]. The electric field truncation, $\Lambda$, determines the dimension of the underlying Hilbert space, $d$, and symmetries of the static charge configuration allow reductions to $d_\mathrm{sym}$. The values of $\Lambda$ are chosen to achieve sub-\% precision in the ground-state energy and representative wavefunctions.  
  }
  \label{tab:SMconfig}
\end{table}

\par In order to calculate the mass of heavy meson, $M_H$, comprised of a static charge (denoted by $Q$ or $\bar{Q}$) and light degrees of freedom,
as well as the two-body and three-body potentials, 17 different configurations of up to three static charges are needed on a four-spatial-site lattice: the empty configuration of the vacuum, a single static charge, five separations of two static charges, and ten three-static-charge configurations.  These configurations are detailed in  Table~\ref{tab:SMconfig}.  In the second column of this table, the symmetric gauge field truncation, $\Lambda$, is chosen to reduce truncation systematic errors on the ground-state energy to below the 1\% precision expected to be attainable with optical quantum hardware.

\par While the systems studied  can be numerically solved with high precision using classical techniques, their dimensionalities are nontrivial with respect to the capabilities of present-day quantum computing devices.  With this in mind, it is convenient to further reduce the latticized, electric-field-truncated Hamiltonians by projecting into the symmetry sectors of momentum $\vec{p}$, parity (P), and charge-parity (CP), as done in Ref.~\cite{Klco:2018kyo}.  
For all but the vacuum state, the presence of static charges at specified lattice points removes the possibility of momentum projecting---the discrete, rotational symmetry of the  lattice has been broken.  Beginning with a Hilbert space dimension of 53, projecting the vacuum to the zero-momentum subspace by creating a basis of rotationally-invariant linear combinations results in a vastly reduced Hilbert space of dimension 15---a size amenable with only four qubits.  Further projecting the vacuum into a sector of positive parity about any one of the four rotationally-equivalent symmetry axes results in a nine-dimensional Hilbert space to be explored with optical hardware.  Such parity projections are possible for eight of the static charge configurations about unique parity axes (e.g., parity axis through sites 2 and 6 for static charges located at 0, 2, 4).

\par Configurations containing one $Q$ and one $\bar{Q}$ do not contain states of definite parity but, rather, of parity and charge conjugation.  As illustrated in Ref.~\cite{Klco:2018kyo}, charge conjugation on a staggered lattice of fermions may be consistently defined by altering the sign of all charges and introducing a shift of a half-spatial-site in order to maintain the staggered distribution of the two-component Dirac spinor with negative/positive charges on even/odd fermion sites. For example, in the case of $Q$ and $\bar{Q}$ at locations 0 and 3, respectively, a parity axis between fermion sites 1 and 5 and a subsequent charge conjugation with clockwise half-spatial-site shift defines a valid CP symmetry
\begin{equation}
C_+P|Q \cdot \cdot \bar{Q} \cdot \cdot \cdot \cdot \rangle = C_+|\cdot \cdot Q \cdot \cdot \cdot \cdot \bar{Q} \rangle = |Q \cdot \cdot \bar{Q} \cdot \cdot \cdot \cdot \rangle.
\end{equation}
Projecting into a sector of positive CP, the sector containing the ground state of the $Q$ and $\bar{Q}$ system, results in a Hilbert space dimensionality reduction from 95 to 58 states. This brings the system within reach of advances in quantum optical devices presented in the main text.

\paragraph*{Potentials and Effective Interactions from Simulations of the Schwinger Model}

Hybrid classical-quantum VQE calculations were performed to determine ground-state energies of systems containing two or three static charges 
from a set of Hamiltonian matrices, providing both eigenvalues and eigenvectors through a customized VQE algorithm using the UCC ansatz.  Differences between these ground-state energies and their wavefunctions reveal the interaction potentials between the static charges and the induced  modifications to the vacuum charge distributions.
To compute the potential energy between static charges located at $r=0$ and $r=3$, for example, the ground-state energy of the Hamiltonian matrix defining the truncated Hilbert space with one charge located at $r=0$ and an anti-charge located at $r=3$, $E^{(Q\overline{Q})}(0,3)$, is determined, along with the wavefunction.
From these the energy of the vacuum is removed to give 
$\Delta E^{(Q\overline{Q})}(3) = E^{(Q\overline{Q})}(0,3) - E^{\rm vac}$.
 This energy difference is independent of where it is evaluated by the discrete rotational symmetry and CP symmetry of the lattice discretization.
A similar calculation is performed of the energy of a single static charge, 
$E^{(Q)}(0)$, that leads to the mass of the heavy meson, $M_H = E^{(Q)}(0) - E^{\rm vac}$. 
The two-body potential between the static charges is defined by the difference
$V^{(Q\overline{Q})}(3) = \Delta E^{(Q\overline{Q})}(3) - 2 M_H$.
The other two-body potentials,
$V^{(QQ)}(r)$ and $V^{(Q\overline{Q})}(r)$ for $r$ even and odd respectively,
are found similarly.
Extraction of the three-body potentials requires a further subtraction, and as an example consider the potential between static charges at $r=0$ and $r=2$ and a static anti-charge at $r=3$.  The energy of the vacuum is subtracted from the ground-state energy,
$\Delta E^{(QQ\overline{Q})}(0,2,3) = E^{(QQ\overline{Q})}(0,2,3) - E^{\rm vac}$.
From this, the masses of three heavy mesons are removed,
$\Delta^2 E^{(QQ\overline{Q})}(0,2,3) = \Delta E^{(QQ\overline{Q})}(0,2,3) - 3 M_H$.
To define the residual three-body potential, the contributions from the two-body interactions are removed,
$V^{(QQ\overline{Q})}(0,2,3) = \Delta^2 E^{(QQ\overline{Q})}(0,2,3) - 
V^{(QQ)}(2)- V^{(Q\overline{Q})}(1) - V^{(Q\overline{Q})}(3)$.
The values obtained in the simulation for the vacuum energy, the mass of the heavy meson, the two-body and three-body potentials, 
obtained with a mass $\mu=0.1$ and hopping term coefficient $x=0.6$ defining the Schwinger model Hamiltonian,
are given in 
Table~\ref{tab:vacMpotentials}.
The two-body and three-body potentials are displayed in 
Fig.~\ref{fig:VrBBexpt}
of the main text and Fig.~\ref{fig:Vr3} appearing later in this section.

\begin{table}[!ht]
\centering
\setlength{\tabcolsep}{15pt}
\renewcommand{\arraystretch}{1.1}
\begin{tabular}{
c
c
r@{$\pm$}l
r@{}l
}
\hline
\hline
$\mathcal{O}$ & $(R_1,R_2)$ & $\braket{\mathcal{O}}$ & $\sigma_{\mathcal{O}}$ & & \hspace{0.05in}exact \\
\hline
\hline
$E_{vac}$ & 	 &$	-2.014 $&$ 0.020	$ &	$-$&$2.0158$	\\
$M_H$ & 	 & $ 	1.283 $&$ 0.021	$ &	& $1.2825$	\\
\hline
$V^{QQ}(0) $ & 	&$ 	0.368 $&$ 0.048	$ &	& $0.3683$	\\
$V^{QQ}(2) $& 	 &$ 	0.136 $&$ 0.048	$ & &	$0.1372$	\\
$V^{QQ}(4) $ & 	 &$ 	0.047 $&$ 0.048	$ & &	$0.0482$	\\
$V^{Q\bar{Q}}(1) $& 	 &$ 	-1.475 $&$ 0.048	$ &	$-$&$1.4756$	\\
$V^{Q\bar{Q}}(3) $& 	 &$ 	-0.262 $&$ 0.047	$ &	$-$&$0.2606$	\\
\hline
$V^{QQQ}(0, 0, 0) $& 	(0, 0) &$	3.49 $&$ 0.17	$ &	& $3.4901$	\\
$V^{QQQ}(0, 0, 2) $& 	(0, 2) &$	-0.05 $&$ 0.13	$ &	$-$&$0.0464$	\\
$V^{QQQ}(0, 0, 4) $& 	(0, 4) &$	0.00 $&$ 0.13	$ &	& $0.0033$	\\
$V^{QQQ}(0, 2, 4) $& 	(2, 3) &$	0.06 $&$ 0.13	$ &	& $0.0561$	\\
$V^{QQ\bar{Q}}(0, 0, 1) $& 	(0, 1) &$	1.29 $&$ 0.13	$ &	& $1.2872$	\\
$V^{QQ\bar{Q}}(0, 0, 3) $& 	(0, 3) &$	0.16 $&$ 0.13	$ &	& $0.1591$	\\
$V^{QQ\bar{Q}}(0, 2, 1) $& 	(2, 0) &$	1.21 $&$ 0.13	$ &	& $1.2097$	\\
$V^{QQ\bar{Q}}(0, 2, 3) $& 	(2, 2) &$	0.14 $&$ 0.11	$ &	& $0.1383$	\\
$V^{QQ\bar{Q}}(0, 2, 5) $& 	(2, 4) &$	0.09 $&$ 0.13	$ &	& $0.0867$	\\
$V^{QQ\bar{Q}}(0, 4, 1) $& 	(4, 1) &$	0.20 $&$ 0.11	$ &	& $0.2008$	\\
\hline
\hline
\end{tabular}
\caption{
The vacuum energy, the mass of the heavy meson, and the two-body and three-body potentials extracted from the VQE implementation.  The uncertainties 
result from propagating $1\%$ systematic uncertainties in the simulated ground-state energies. The second column indicates the Jacobi coordinates for the three-body systems. The final column shows the calculated values of the potentials at the simulation-implemented values of $\Lambda$, electric field truncation.
}
\label{tab:vacMpotentials}
\end{table}

The ground-state energies determined with VQE have both statistical uncertainties, determined by variations in the last several iterations, and systematic uncertainties.  The dominant systematic uncertainty is reproducibility of the simulation results, which was estimated by variations in results collected during
multiple long runs on a representative set of Hamiltonian matrices, repeated throughout the course of the data collection.  This variation was found to be less than one percent, and we assign a systematic uncertainty of $1\%$ to each energy measurement as a conservative estimate.

Beyond numerical determination of the two-body potentials between static charges, it is worth making the connection to nuclear physics phenomenology through parameterization of the potentials based upon the spectrum of the Schwinger model, and through matching to the appropriate low-energy EFT.
In the 1+1 dimensional Schwinger model, 
the potential between charges falls with distance as the sum of exponentials, as the spectrum does not contain a massless particle.
With the parameters that were used in the simulation, the number of measurements of the potentials are few, three for the $QQ$ systems and two for the $Q\overline{Q}$ system.  Consequently, we fit a single exponential in both channels, with the understanding that they are expected to reproduce the correct behaviors at long distances, but are merely parameterizations at intermediate and short distances.
Results obtained for, and from, these parameterizations have associated unquantified model  uncertainties.
We write the parametrizations of the 
infinite-volume
two-body potentials as 
\begin{eqnarray}
V^{(QQ)}(r) & = & 
\left(g^{(QQ)}\right)^2\ e^{-M^{(QQ)} r}
\ \ ,\ \ 
V^{(Q\overline{Q})}(r) \ =\  
-\left(g^{(Q\overline{Q})}\right)^2\ e^{-M^{(Q\overline{Q})} r}
\ \ \ ,
\label{eq:sup:pots}
\end{eqnarray}
where the couplings,
$g^{(QQ)}$ and $g^{(Q\overline{Q})}$, and the masses
$M^{(QQ)}$ and $M^{(Q\overline{Q})}$,
are treated as fit parameters.
We expect the masses to be close to the mass of the lightest vector meson, but modified by the close proximity of other states.

As our calculations are performed in a finite volume subject to periodic boundary conditions, the potentials experienced by  static charges are modified by the presence of image charges, separated by a distance $n L$, where $n$ is an integer and $L$ is the spatial extent of the lattice.  As a result, the potentials extracted from our lattice calculations will be of the form~\cite{Luscher:1985dn,Luscher:1986pf},
\begin{eqnarray}
V^{(j),L}(r) & = & 
\sum_{n=-\infty}^{+\infty}
\ V^{(j)}(|r+ n L|) 
\ \ \ ,
\label{eq:sup:periodicV}
\end{eqnarray}
where $j=QQ, Q\overline{Q}$ correspond to the potentials in Eq.~(\ref{eq:sup:pots}).
\begin{figure}
    \centering
    \includegraphics[height=0.29\textheight]{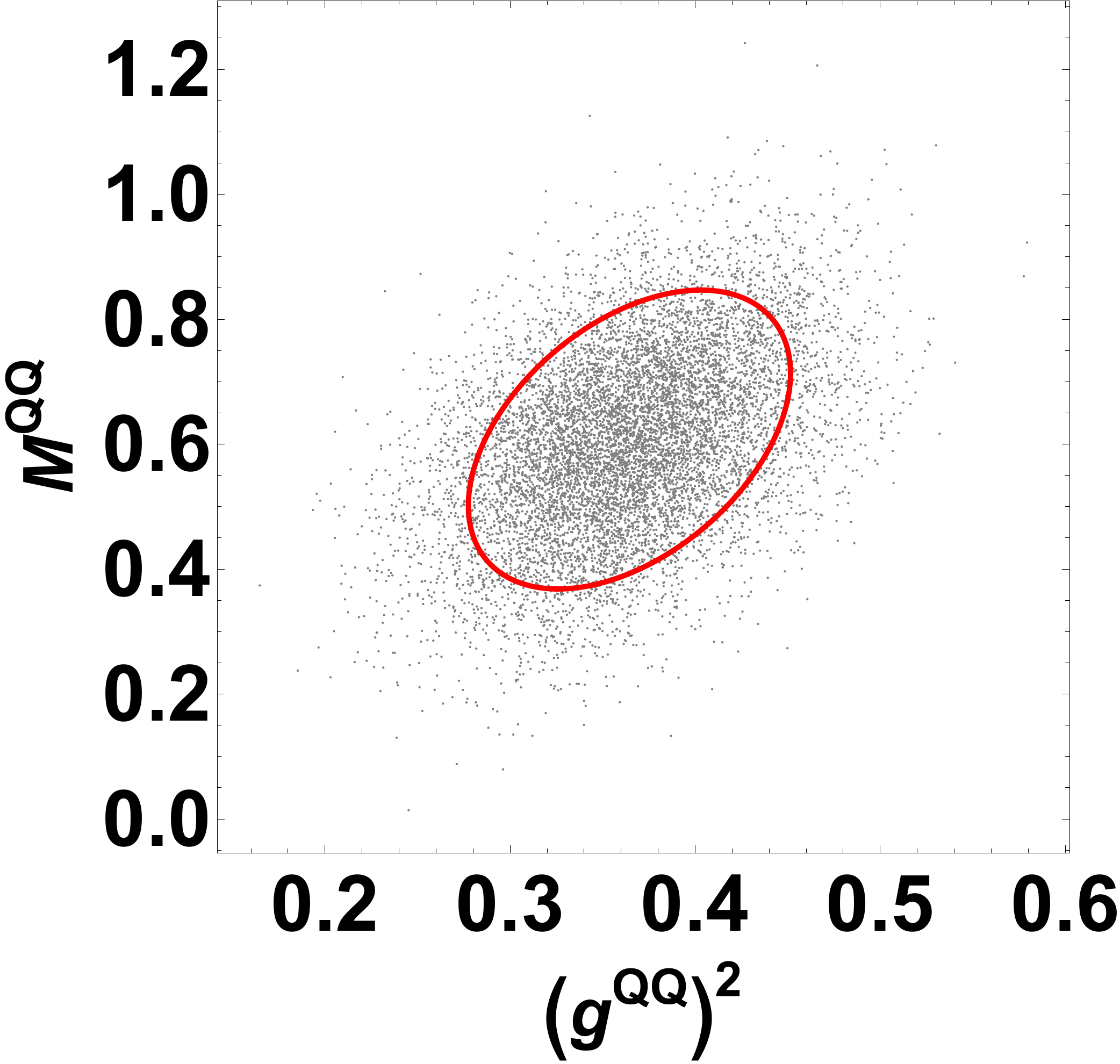}
    \includegraphics[height = 0.29\textheight]{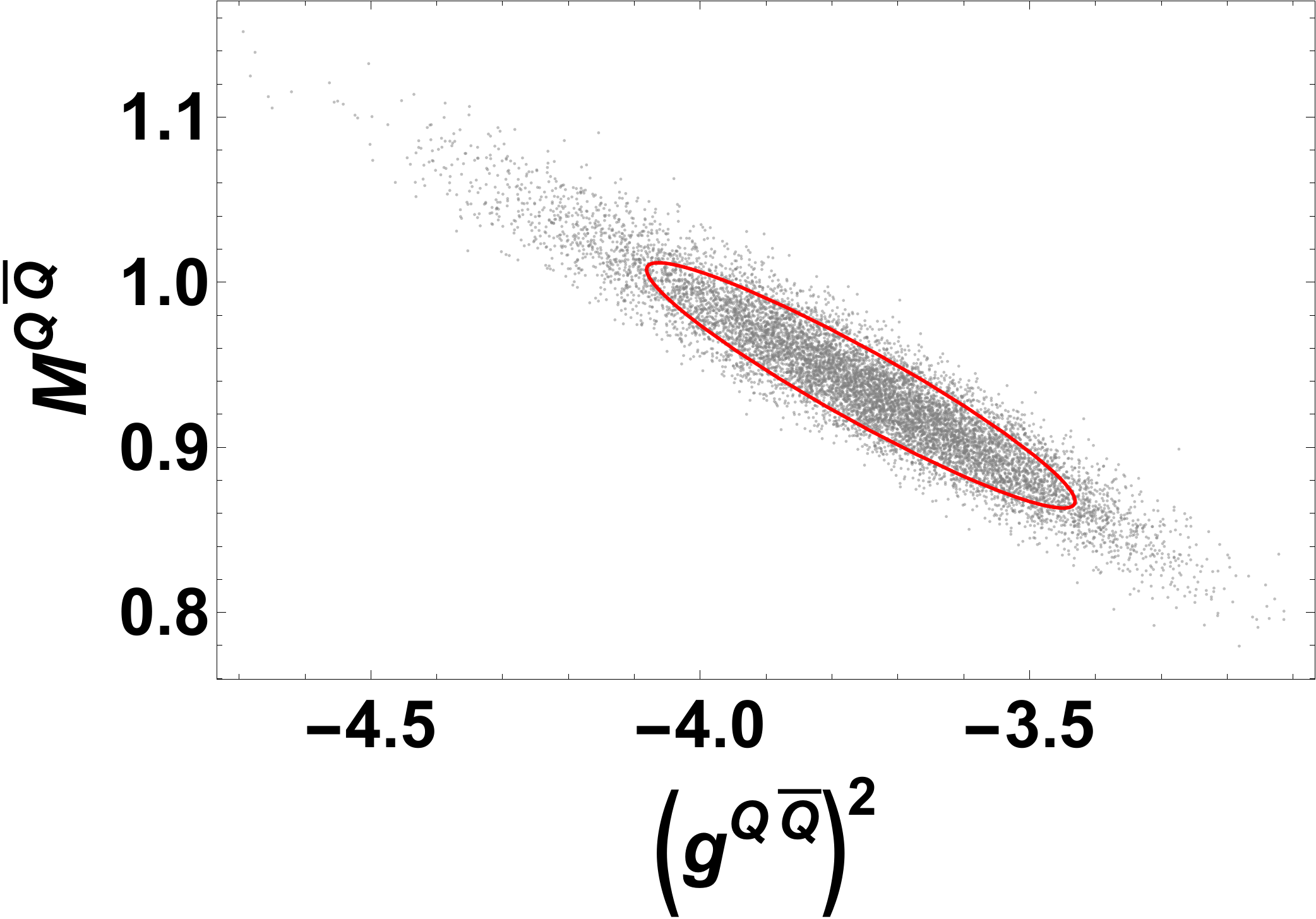}
    \caption{Monte Carlo samples and 68\% confidence ellipses of the couplings and masses of the $QQ$ and $Q\overline{Q}$ two-body interactions. The ellipse associated with the $QQ$ system (left panel) is described by eigenvectors $ (0.176, 0.984)$ and $  (-0.984, 0.176)$ with associated semi-axis radii 0.177 and 0.984.  The ellipse associated with the $Q\overline{Q}$ system (right panel) is described by eigenvectors $(-0.977, 0.211)$ and $(-0.211, -0.977)$ with associated semi-axis radii 0.332 and 0.0236. }
    \label{fig:ellipses}
\end{figure}
Fitting these forms for the two-body potentials 
to the results obtained with the QFP
leads to 
\begin{eqnarray}
\left(g^{(QQ)}\right)^2 & = & 0.365(52)
\ \ ,\ \ 
M^{(QQ)}\ =\ 0.61(14)
\nonumber\\
\left( g^{(Q\overline{Q})} \right)^2 & = & 
3.75(31)
\ \ ,\ \ 
M^{(Q\overline{Q})}\ =\ 
0.937(71)
\ \ \ ,
\label{eq:sup:2bofyfits}
\end{eqnarray}
where the quoted uncertainties are determined by projection of the elliptical contours of Fig.~\ref{fig:ellipses} onto each axis---resulting uncertainties being slightly enlarged with respect to those quoted for single-variable,  marginalized probability distributions.
These quantities have support in the ultraviolet structure of the theory, and are modified in the finite volume by terms that are  exponentially small, and determined by the ratio of $\Lambda L$, where $\Lambda$ is the ultraviolet scale~\cite{Luscher:1985dn,Luscher:1986pf}.
Assuming that there is a scale separation between the masses in the finite-volume effects and the ultraviolet scale, the fit parameters in the potentials can be used to extrapolate to infinite volume simply by inserting them into the potentials in
Eq.~(\ref{eq:sup:pots}).  Both the periodic and infinite volume potentials are shown in the left panel of Fig.~\ref{fig:VrBBexpt} of the main text.

These potentials can be used directly for phenomenological applications (in 1+1 dimensions) for processes
involving heavy mesons of finite mass and momentum up to the ultraviolet scale of the theory.  Results from these potentials are expected to exhibit deviations from actual predictions of the Schwinger model due to the limited form fit to the data.  For low-energy processes, calculations can be reorganized and generally made simpler by matching to a low-energy EFT with consistent power counting that is explicitly constructed to faithfully reproduce the low-energy behavior of S-matrix elements.
In nuclear physics, the low-energy behavior (below the t-channel cut in one-pion exchange) of few nucleon systems is reproduced by the pionless EFT~\cite{Kaplan:1998we,Kaplan:1998tg,vanKolck:1998bw}, consisting of contact operators of delta-functions and (covariant-)derivatives, and gauge-invariant operators describing interactions with external probes~\cite{Chen:1999tn}.

As an example, we outline the matching between the Schwinger model and its low-energy EFT, in which there are only dynamical heavy mesons.  
Using the fit values of the EFT parameters and their associated uncertainties in the infinite-volume $Q\overline{Q}$ potential, 
we solve the Schr\"{o}dinger equation to produce zero-energy wavefunctions.  Far from the potential, the wavefunctions are straight lines, and define the scattering lengths.
These wavefunctions can be reproduced by a delta-function potential with strength $C_0$ arising from an effective potential of the form  $V_{\rm eff}(r) = C_0 \delta(r)$.
For a heavy meson of mass $M_H^{(EFT)}=4.5$ that is chosen for the sake of demonstration only, $C_0 = -0.117(30)$ which should be compared to the naive estimate from the Born term of $C_0 = -8.0(0.2)$.
The potential also admits two bound states,
a ``shallow'' one with $E\sim -0.59$ and a deep one with $E\sim-2.4$, which corresponds to a positronium-like state.

It is convenient to work with Jacobi coordinates in presenting the three-body potentials.  These relative coordinates are defined by 
$R_1 = |r_1-r_2|$ and $R_2 = r_3 - R_1/2$,
where we have worked with the convention that the first two particles are identified as those with the same charge.  The results of our experiments are presented in 
Table~\ref{tab:vacMpotentials}.
and displayed in 
Fig.~\ref{fig:Vr3}.
\begin{figure}[!ht]
    \centering
    \includegraphics[width = 0.85\textwidth]{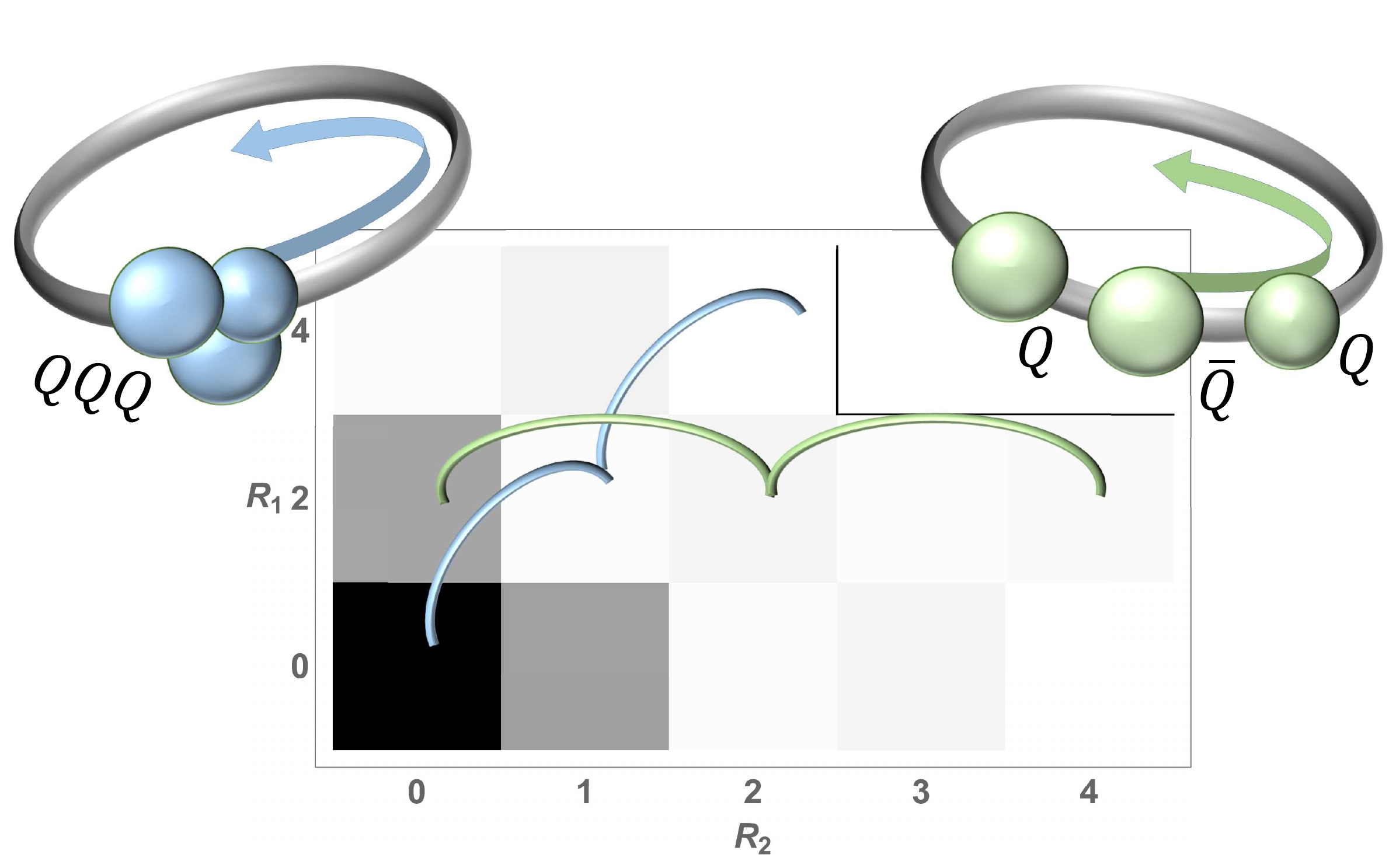}
    \caption{The potential between three static charges represented by Jacobi coordinates in 1-dimension, $R_1 =| r_2 - r_1|$ and $R_2 = \big|r_3 - \frac{R_1}{2} \big|$ with $r_{1,2,3}$ the $QQQ$ or $QQ\overline{Q}$ distances from the origin.  The physical configurations of static charges on the lattice associated with the blue and green paths through the grid of three-body potential values are depicted by the schematic diagrams at the corners. 
    }
    \label{fig:Vr3}
\end{figure}
The three-body potentials are found to fall rapidly with either of the Jacobi coordinates, as expected.   
While these potentials could be matched to the low-energy EFT, with operator structures of the form
${\cal O}\sim \left(N^\dagger N\right)^3$, to be used in other more complex calculations, we leave that for future investigations.
The deformations to the vacuum structure resulting from these three-body forces, arrived at by taking differences in the energy density in the electric field and in the probabilities of the electron and positron states, have been calculated.  
In
Fig~\ref{fig:r021local}, we show the modifications to the vacuum structure for 
$V^{(QQ\overline{Q})}(0,2,1)$, corresponding to the three-body system with Jacobi coordinates $R_1=2$ and $R_2=1$.
\begin{figure}[!ht]
    \centering
    \includegraphics[width = 0.45\textwidth]{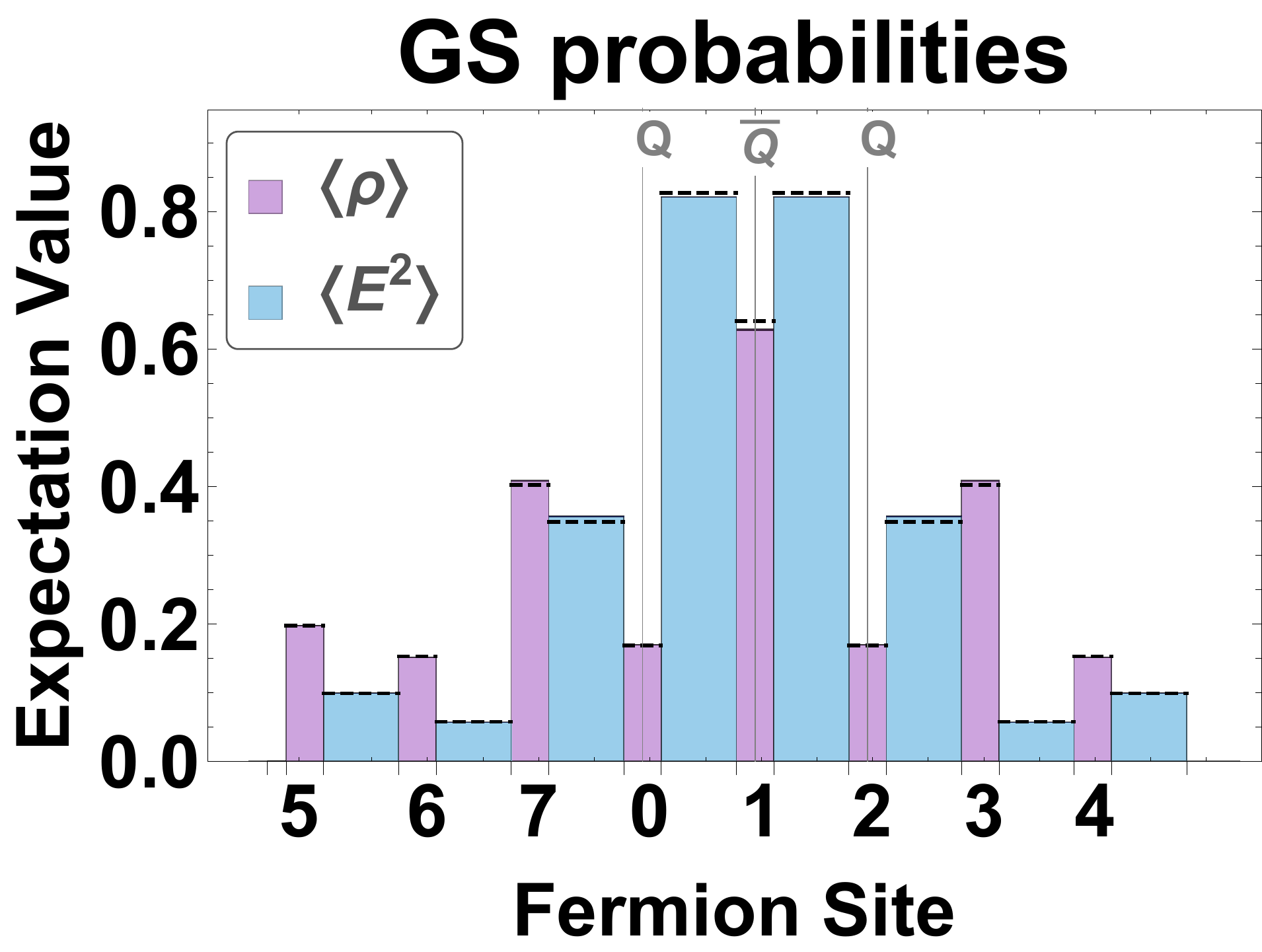}
    \includegraphics[width=0.45\textwidth]{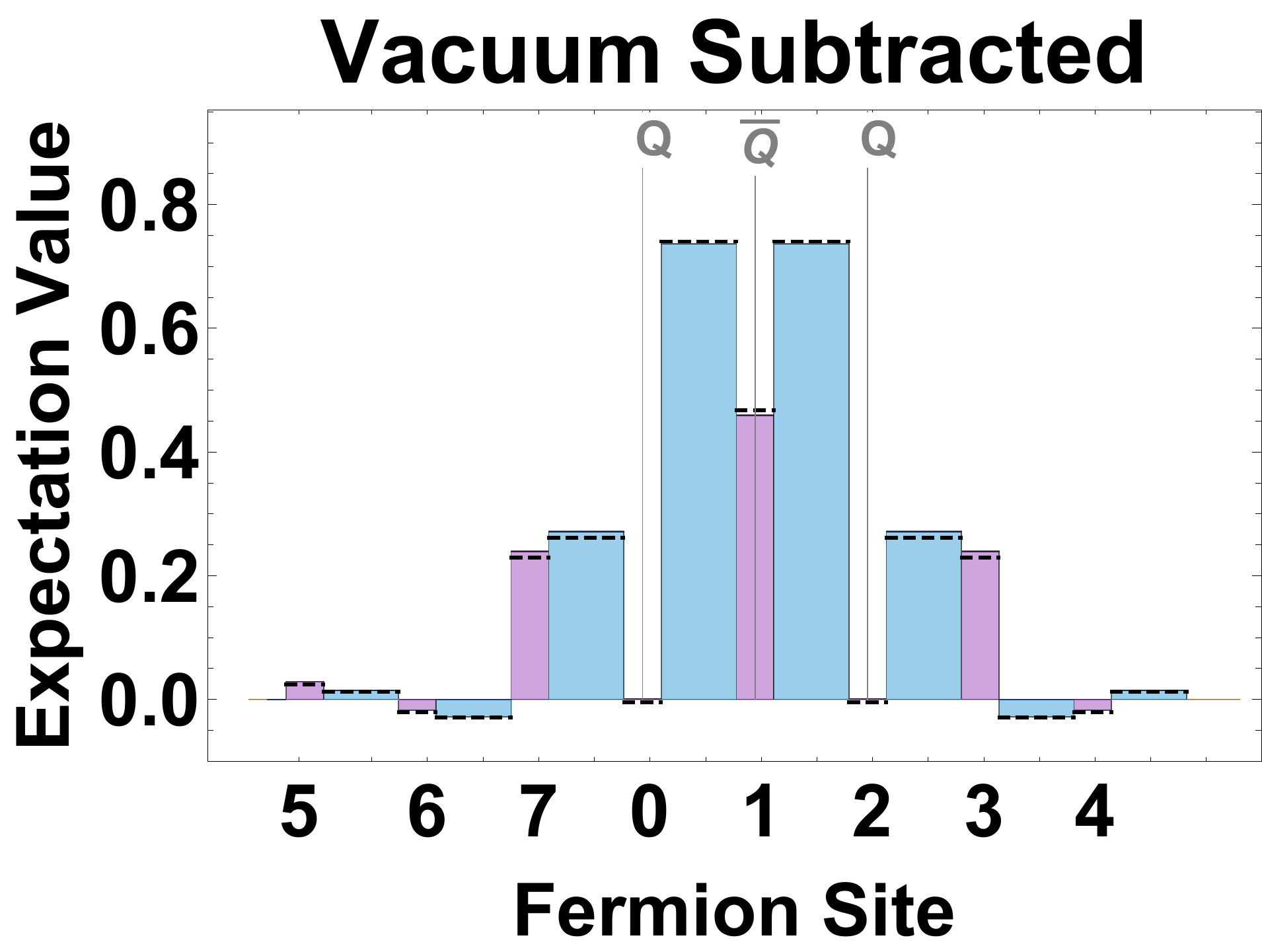}
    \includegraphics[width = 0.45\textwidth]{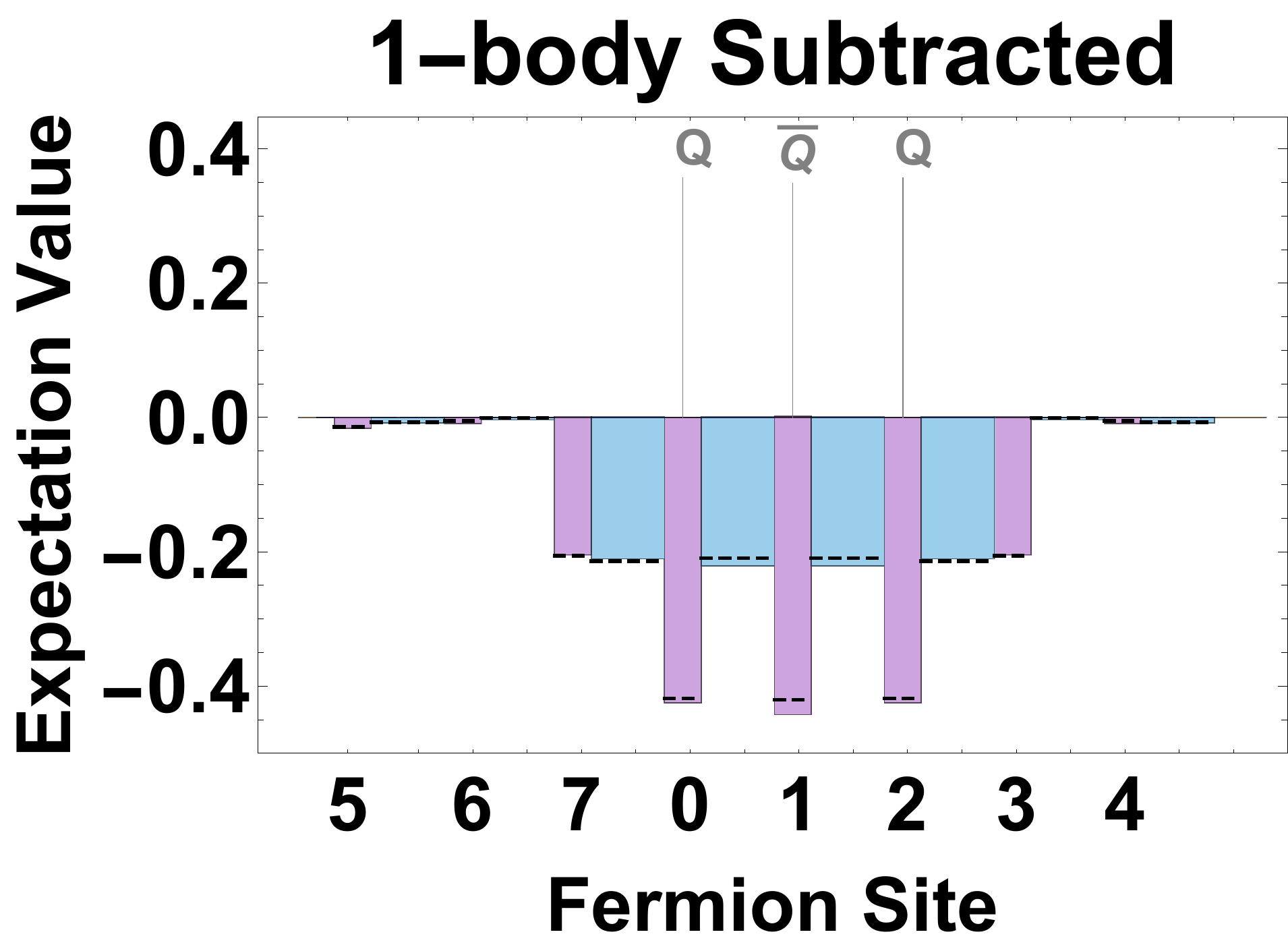}
    \includegraphics[width = 0.45\textwidth]{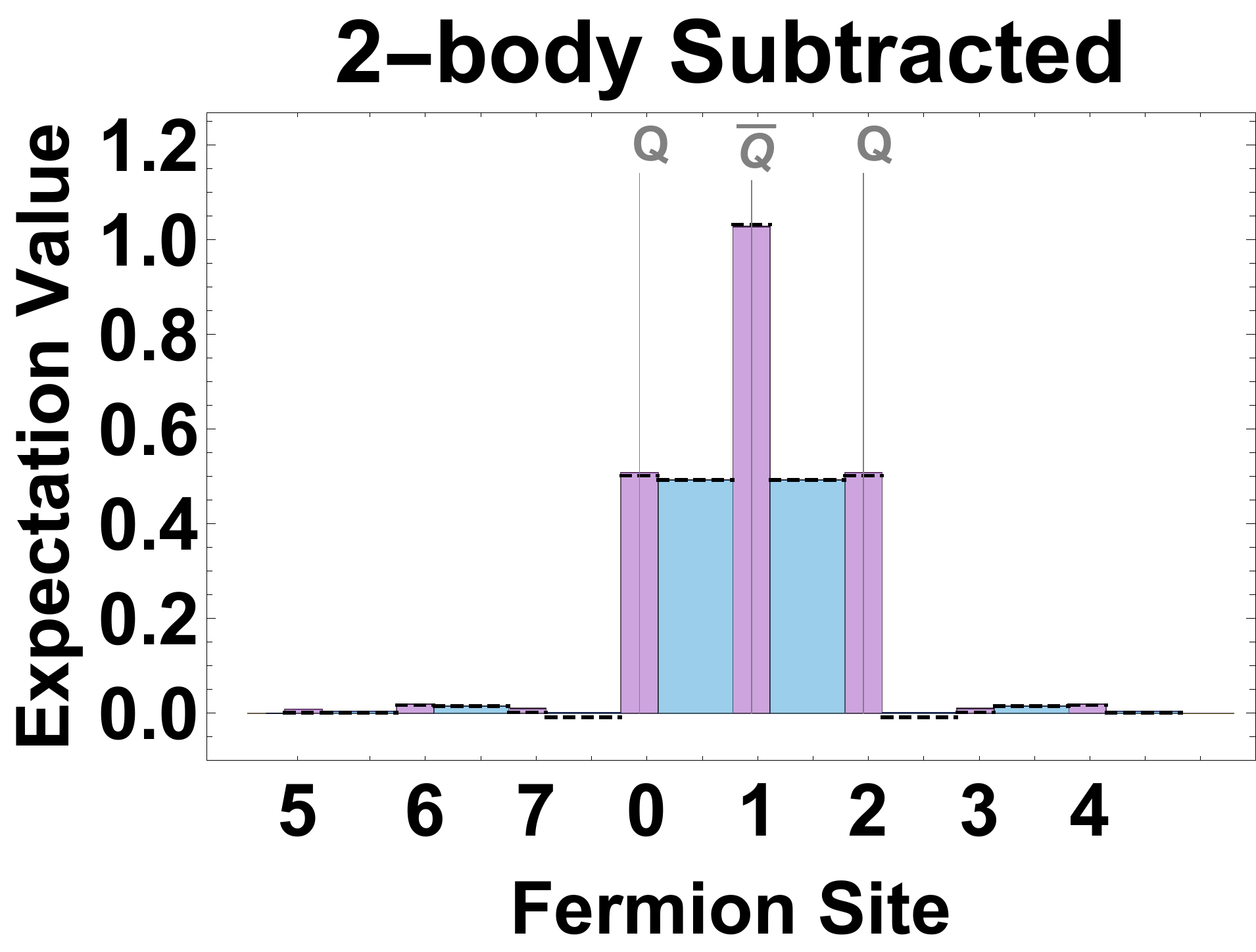}
    \caption{Local properties of the vacuum with three static charges, $Q$ and $\bar{Q}$ located at lattice sites (0, 2) and 1, respectively.  
    The uncertainties, which are too small to be visible, represent the stability of these local properties over wavefunctions extracted over the last ten VQE iterations.  The horizontal dashed lines are the values calculated through exact diagonalization. The values shown in each panel are presented in Table~\ref{tab:localr3}.
    }
    \label{fig:r021local}
\end{figure}

A single-hadron quantity that we derive from our results is the charge radius of the heavy meson.  Unlike classical lattice QCD calculations, where contributions from both connected-quark and disconnected-quark must be calculated, unless a symmetry forbids one or both of the contributions, the quantum computation allows for a direct determination of relevant quantities from the wavefunction of the system.  In the case of the heavy meson formed around a static quark at $r=0$, the charge radius can be determined by a direct evaluation of the discrete sum
\begin{eqnarray}
\langle r^2 \rangle_Q
& = & 
\sum_{n=0}^{N_Q/2}\ 
(-1)^n\ n^2\ {\rm Prob}(n) 
\ \ \  ,
\end{eqnarray}
where ${\rm Prob}(n)$ is the probability of finding an electron or positron at the $n^{\rm th}$ site.  The sum is cut off at half of the lattice to minimize the contribution from the image charges, introducing an uncertainty
naively estimated to be the average size of the last two contributions.
\begin{figure}[!ht]
    \centering
    \includegraphics[width=0.32\textwidth]{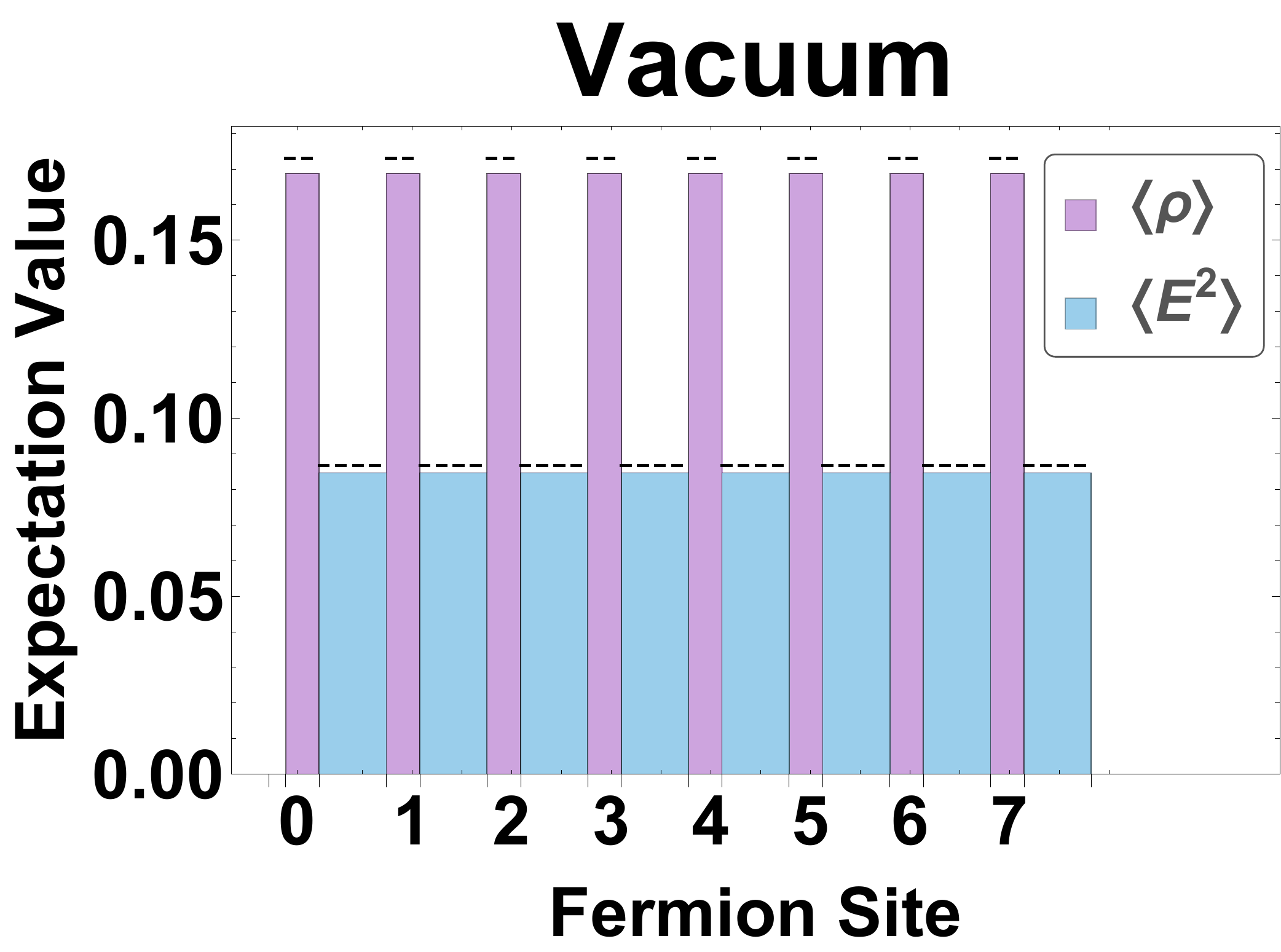}
    \includegraphics[width = 0.32\textwidth]{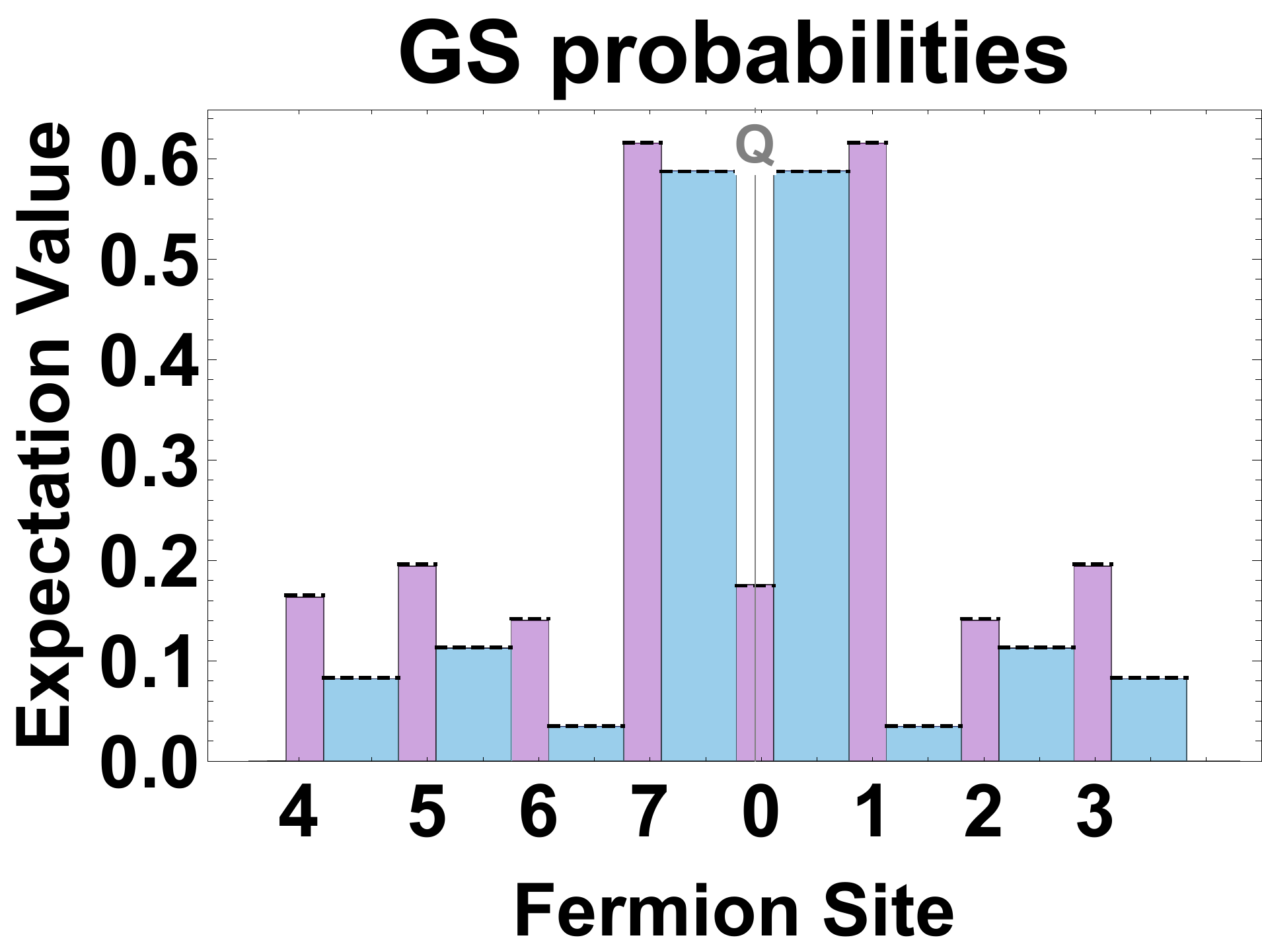}
    \includegraphics[width = 0.32\textwidth]{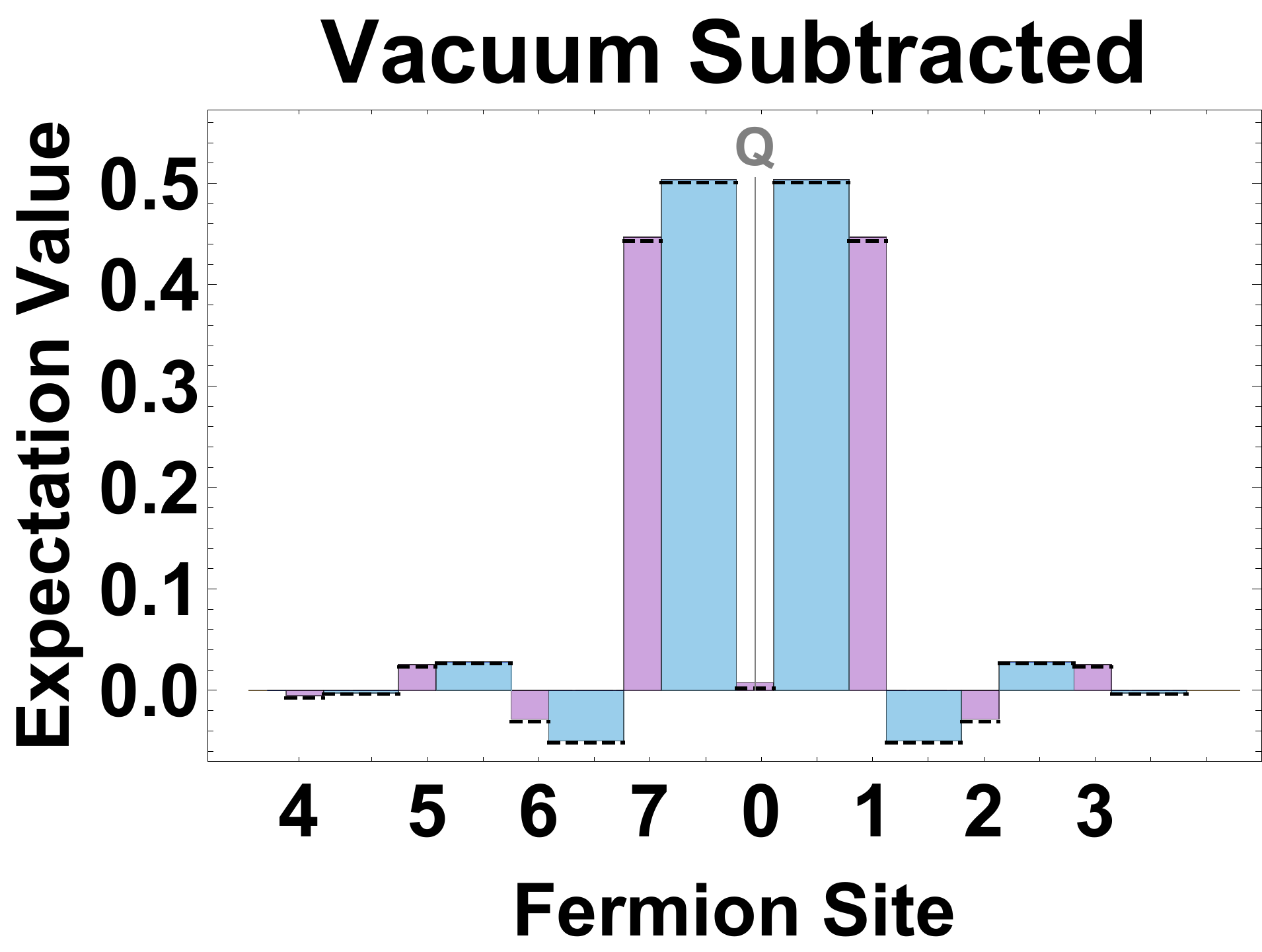}
    \caption{Local properties of the vacuum and a single static charge located at $r=0$ that creates a heavy meson.  The uncertainties, which are too small to be visible, represent the stability of these local properties over wavefunctions extracted over the last ten VQE iterations.  The horizontal dashed lines are the values calculated through exact diagonalization.
    The values shown in each panel are presented in Table~\ref{tab:vacuum} and Table~\ref{tab:r0}.}
    \label{fig:vacuumAND1Blocal}
\end{figure}
We find that the square-charge radius of the heavy meson containing a positively charged static charge,
determined from the charge distribution shown in Fig.~\ref{fig:vacuumAND1Blocal},
to be
\begin{eqnarray}
\langle r^2 \rangle_Q
& = & 
-1.76(32)
\ \ \ ,
\end{eqnarray}
in lattice units.
Similarly, the radius of the energy density in the electric field can be computed,
\begin{eqnarray}
\langle r^2 \rangle_{E^2}
& = & 
0.33(15)
\ \ \ .
\end{eqnarray}
While there appears to be a large difference between these two radii, one must keep in mind that they are derived by weighting with distances that are constrained by the lattice spacing, or half lattice spacing, and presently  unquantified discretization effects are expected to be significant in these quantities.  Further calculations at a smaller lattice spacing are required to perform an extrapolation to the continuum limit and to provide  a complete quantification of uncertainties.

\par It is interesting to note that there are static-charge configurations, such as the three-body $QQ\bar{Q}$ system at locations (0, 4, 1), that do not allow reduction of the Hilbert space through symmetry projections.  It is this system in particular that has required the largest Hilbert space (68 dimensions) to achieve 1\% precision on the ground-state energy used in the calculation of the corresponding three-body potential.   While these symmetry projections have been critical for constructing systems with a dimension manageable with current quantum hardware and for removing dynamically-irrelevant sectors from the perspective of the ground-state properties, it is interesting to note that knowledge of the symmetry properties without explicit projection could be used to probe systematic errors or noise within the quantum computation.  For example, the local expectation values of the charge density and energy in the electric field shown in Fig.~\ref{fig:r021local} for the (0, 2, 1) $QQ\overline{Q}$ system are currently forced theoretically to satisfy the parity projection through fermion site 1.  If the 67-dimensional system (that before parity projection) was instead implemented by the QFP, deviations from this exact spatial symmetry would be indicative of the systematic uncertainty in the structure of the ground-state wavefunction.

\paragraph*{Calculated values}

In this section, we present the values of local probabilities in the ground state (GS) of the vacuum, a single static charge, and two- and three-static charge systems, that are shown in Figs.~\ref{fig:r021local}, \ref{fig:vacuumAND1Blocal} and Fig.~\ref{fig:VBBr2dist}.
\begin{table}[!ht]
    \centering
    GS probabilities associated with two static charges, at $r=0,2$ \\
    \begin{tabular}{c|cc|cc}
    \hline
    \hline
    $r$ & $\langle \rho \rangle $ & $\langle \rho \rangle_\text{exact} $ & $\langle E^2 \rangle$ & $\langle E^2 \rangle_\text{exact}$ \\
    \hline
    0	&	0.115596(19) &	0.115881	&	0.29668(15)	&	0.297446 \\
    1	&	0.59336(29)	 &	0.594892	&	0.29668(15)	&	0.297446\\
    2	&	0.115596(19) &	0.115881	&	0.81892(16)	&	0.818435 \\
    3	&	0.82624(16)	 &	0.826473	&	0.018040(24)&	0.01883 \\
    4	&	0.121542(27) &	0.125071	&	0.114283(24)&	0.117114 \\
    5	&	0.228433(50) &	0.234066	&	0.114283(35)&	0.117114 \\
    6	&	0.121542(27) &	0.125071	&	0.018040(25)&	0.01883 \\
    7	&	0.82624(16)	&	0.826473	&	0.81892(16)	&	0.818435 \\
    \hline
    \hline
    \end{tabular}
    \\\vspace{0.2cm} Vacuum Subtracted \vspace{0.2cm} \\
    \begin{tabular}{c|cc|cc}
    \hline
    \hline
    0	&	$-$0.053082(14)	&	$-$0.0571012	&	0.21208(15) &	0.210749 \\
    1	&	0.42468(30) &	0.42191	&	0.21208(15)	& 0.210749 \\
    2	&	$-$0.053082(14)	&	$-$0.0571012	&	0.73432(16) &	0.731738 \\
    3	&	0.65757(15) &	0.653491	&	$-$0.066560(31)	&	$-$0.0678671 \\
    4	&	$-$0.047135(27)	&	$-$0.0479109	&	0.029682(21)	&	0.0304174 \\
    5	&	0.059756(43)	&	0.0610843	&	0.029682(21)	&	0.0304174 \\
    6	&	$-$0.047135(27)	&	$-$0.0479109	&	$-$0.066560(31)	&	$-$0.0678671 \\
    7	&	0.65757(15)	&	0.653491	&	0.73432(16)	&	0.731738 \\
\hline
\hline
    \end{tabular}
    \\\vspace{0.2cm} 1-body Subtracted \vspace{0.2cm} \\
    \begin{tabular}{c|cc|cc}
    \hline
    \hline
    0	&	$-$0.031897(61)	&	$-$0.0278071	&	$-$0.24139(13)	&	$-$0.238353 \\
1	&	$-$0.46919(27)	&	$-$0.46415	&	$-$0.24139(13)		&	$-$0.238353 \\
2	&	$-$0.031897(61)	&	$-$0.0278071	&	0.20279(19)	&	0.204377 \\
3	&	0.18521(19)	&	0.187407	&	$-$0.0140797(97)	&	$-$0.012637 \\
4	&	$-$0.013040(63)	&	$-$0.00937306	&	0.004098(51)	&	0.00738809 \\
5	&	0.00891(10)	&	0.0149761	&	0.004098(51)	&	0.00738809 \\
6	&	$-$0.013040(63)	&	$-$0.00937306	&	$-$0.0140797(97)	&	$-$0.012637 \\
7	&	0.18521(19)	&	0.187407	&	0.20279(19)	&	0.204377 \\
\hline
\hline
    \end{tabular}
    \caption{Measured and exact expectation values for the local charge density and energy in the electric field as shown in Fig.~\ref{fig:VBBr2dist}  of the main text with two static charges located at sites zero and two.  The uncertainties on measured values represent statistical fluctuations over the last ten wavefunctions of the VQE iterations and do not include estimates of associated  systematic uncertainties. Note, from Table~\ref{tab:SMconfig}, that parity has been enforced for this system, leading to spatially symmetric locations having the same values.}
    \label{tab:localr2}
\end{table}
Table~\ref{tab:localr2} contains the 
measured and exact expectation values of the local charge density and energy in the electric field with two static charges located at $r=0, 2$, as shown in Fig.~\ref{fig:VBBr2dist}  of the main text.
\begin{table}[!ht]
\vspace{-2cm}
    \centering
    GS probabilities associated with three static charges, at $r=0,1,2$\\
    \begin{tabular}{c|cc|cc}
    \hline
    \hline
    $r$ & $\langle \rho \rangle $ & $\langle \rho \rangle_\text{exact} $ & $\langle E^2 \rangle$ & $\langle E^2 \rangle_\text{exact}$ \\
    \hline
    0	&	0.16927(16)	&	0.168397	&	0.82064(80) &	0.826857 \\
    1	&	0.6272(17)	&	0.640306	&	0.82064(80)	&	0.826857 \\
    2	&	0.16927(16)	&	0.168397	&	0.35567(99)	&	0.348244 \\
    3	&	0.40786(86)	&	0.402136	&	0.05647(12)	&	0.057447 \\
    4	&	0.150959(62)	&	0.152525	&	0.099227(72)	&	0.0989282 \\
    5	&	0.19755(14)	&	0.197266	&	0.099227(72)&	0.0989282 \\
    6	&	0.150959(62)	&	0.152525	&	0.05647(12)	&	0.057447 \\
    7	&	0.40786(86)	&	0.402136	&	0.35567(99)	&	0.348244 \\
    \hline
    \hline
    \end{tabular}
     \\\vspace{0.2cm} Vacuum Subtracted \vspace{0.2cm} \\
    \begin{tabular}{c|cc|cc}
    \hline
    \hline
    0	&	0.00060(15)		&	$-$0.00458463	&	0.73604(81)	&	0.74016 \\
    1	&	0.4585(17)	&	0.467324	&	0.73604(81)	&	0.74016 \\
    2	&	0.00060(15)	&	$-$0.00458463	&	0.27107(99)	&	0.261547 \\
3	&	0.23918(85)	&	0.229154	&	$-$0.02813(12)	&	$-$0.0292501 \\
4	&	$-$0.017718(72)	&	$-$0.0204571	&	0.014626(67)	&	0.0122311 \\
5	&	0.02887(14)	&	0.0242844	&	0.014626(67)	&	0.0122311 \\
6	&	$-$0.017718(72)	&	$-$0.0204571	&	$-$0.02813(12)	&	$-$0.0292501 \\
7	&	0.23918(85)	&	0.229154	&	0.27107(99)	&	0.261547 \\
\hline
\hline
    \end{tabular}
    \\\vspace{0.2cm} 1-body Subtracted \vspace{0.2cm} \\
    \begin{tabular}{c|cc|cc}
    \hline
    \hline
    0	&	$-$0.42515(22) &	$-$0.418321	&	$-$0.22082(77)	&	$-$0.209597 \\
    1	&	$-$0.4427(16)	&	$-$0.420459	&	$-$0.22082(77)	&	$-$0.209597 \\
    2	&	$-$0.42515(22)	&	$-$0.418321	&	$-$0.2105(10)	&	$-$0.21426 \\
    3	&	$-$0.20467(92)	&	$-$0.205913	&	$-$0.003783(77)	&	$-$0.000725653 \\
    4	&	$-$0.009048(21)	&	$-$0.00497328	&	$-$0.00841(12)	&	$-$0.00712178 \\
    5	&	$-$0.01640(23)	&	$-$0.014303	&	$-$0.00841(12)	&	$-$0.00712178 \\
    6	&	$-$0.009048(21)	&	$-$0.00497328	&	$-$0.003783(77)	&	$-$0.000725653 \\
    7	&	$-$0.20467(92)	&	$-$0.205913	&	$-$0.2105(10)	&	$-$0.21426 \\
    \hline
    \hline
    \end{tabular}
    \\\vspace{0.2cm} 2-body Subtracted \vspace{0.2cm} \\
    \begin{tabular}{c|cc|cc}
    \hline
    \hline
    0	&	0.50834(19)	&	0.501635	&	0.49244(61)	&	0.492946	\\
1	&	1.0275(13)	&	1.03178	&	0.49244(61)	&	0.492946	\\
2	&	0.50834(19) &	0.501635	&	0.00088(82)	&	$-$0.00837237	\\
3	&	0.00966(69)	&	0.00184607	&	0.01489(11)	&	0.0147168	\\
4	&	0.01936(10)	&	0.0168098	&	0.004382(45) &	0.00071903	\\
5	&	0.008576(91)	&	0.00141658	&	0.004382(45)&	0.00071903	\\
6	&	0.01936(10)	&	0.0168098	&	0.01489(11)	&	0.0147168	\\
7	&	0.00966(69)	&	0.00184607	&	0.00088(82)	&	$-$0.00837237	\\
\hline
\hline
\end{tabular}
    \caption{Measured and exact expectation values for the local charge density and energy in the electric field as shown in Fig.~\ref{fig:r021local} with three static charges located at sites $r=0,1,2$.  The uncertainties on measured values represent statistical fluctuations over the last ten wavefunctions of the VQE iterations and do not include estimates of associated  systematic uncertainties. Note, from Table~\ref{tab:SMconfig}, that parity has been enforced for this system, leading to spatially symmetric locations having the same values.}
    \label{tab:localr3}
\end{table}
Table~\ref{tab:localr3} contains the 
same quantities for three static charges located at 
$r=0,1,2$  as shown in Fig.~\ref{fig:r021local} .
\begin{table}[!ht]
    \centering
   GS probabilities associated with Vacuum \\
    \begin{tabular}{c|cc|cc}
    \hline
    \hline
    $r$ & $\langle \rho \rangle $ & $\langle \rho \rangle_\text{exact} $ & $\langle E^2 \rangle$ & $\langle E^2 \rangle_\text{exact}$ \\
    \hline
    0-7	&	0.168677(14)	&	0.172982	&	0.0846006(74)	&	0.0866971 \\
    \hline
    \hline
\end{tabular}
\caption{Measured and exact expectation values for the vacuum local charge density and energy in the electric field as shown in Fig.~\ref{fig:vacuumAND1Blocal} (with zero external static charges).  The uncertainties on measured values represent statistical fluctuations over the last ten wavefunctions of the VQE iterations and do not include estimates of associated  systematic uncertainties. Note, from Table~\ref{tab:SMconfig}, that parity and translation invariance have been enforced.}
\label{tab:vacuum}
\end{table}
Table~\ref{tab:vacuum} contains the 
measured and exact expectation values for the vacuum local charge density and energy in the electric field as shown in Fig.~\ref{fig:vacuumAND1Blocal}.
\begin{table}[!ht]
    \centering
   GS probabilities associated with one static charge, at $r=0$ \\
    \begin{tabular}{c|cc|cc}
    \hline
    \hline
    $r$ & $\langle \rho \rangle $ & $\langle \rho \rangle_\text{exact} $ & $\langle E^2 \rangle$ & $\langle E^2 \rangle_\text{exact}$ \\
    \hline
    0	&	0.176000(24)	&	0.174705	&	0.588000(12)&	0.587352 \\
1	&	0.615612(14)	&	0.616012	&	0.0346655(84)	&	0.0351434 \\
2	&	0.140169(14)	&	0.141965	&	0.112730(12)	&	0.113403 \\
3	&	0.194102(24)	&	0.196036	&	0.082055(12)	&	0.0830207 \\
4	&	0.163090(23)		&	0.165461	&	0.082055(12)	&	0.0830207 \\
5	&	0.194102(24)	&	0.196036	&	0.112730(12)	&	0.113403 \\ 
6	&	0.140169(14)		&	0.141965	&	0.0346655(84)	&	0.0351434 \\ 
7	&	0.615612(14)	&	0.616012	&	0.588000(12)	&	0.587352 \\
    \hline
    \hline
    \end{tabular}
    \\\vspace{0.2cm} Vacuum Subtracted \vspace{0.2cm} \\
    \begin{tabular}{c|cc|cc}
    \hline
    \hline
    0	&	0.007323(37)	&	0.00172308	&	0.503399(19)	&	0.500655 \\
1	&	0.446935(26)	&	0.44303	&	$-$0.049935(15)	&	$-$0.0515537 \\ 
2	&	$-$0.028508(25)	&	$-$0.0310171	&	0.028130(18)	&	0.0267057 \\ 
3	&	0.025425(33)	&	0.0230541	&	$-$0.002545(17)	&	$-$0.0036764 \\ 
4	&	$-$0.005587(30)	&	$-$0.00752077	&	$-$0.002545(17)	&	$-$0.0036764 \\ 
5	&	0.025425(33)	&	0.0230541	&	0.028130(18)	&	0.0267057 \\ 
6	&	$-$0.028508(25)	&	$-$0.0310171	&	$-$0.049935(15)	&	$-$0.0515537 \\ 
7	&	0.446935(26)	&	0.44303	&	0.503399(19)	&	0.500655 \\ 
\hline
\hline
\end{tabular}
\caption{Measured and exact expectation values for the local charge density and energy in the electric field as shown in Fig.~\ref{fig:vacuumAND1Blocal} with one static charge located at site zero.  The uncertainties on measured values represent statistical fluctuations over the last ten wavefunctions of the VQE iterations and do not include estimates of associated  systematic uncertainties. Note, from Table~\ref{tab:SMconfig}, that parity has been enforced for this system, leading to spatially symmetric locations having the same values.}
\label{tab:r0}
\end{table}
Table~\ref{tab:r0} contains the same quantities for one static charge located at $r=0$, as shown in Fig.~\ref{fig:vacuumAND1Blocal}.

\end{document}